\documentclass[titlepage, reqno, 12pt]{amsart}
\usepackage{url, soul, amssymb, color, amsmath, graphicx, amsaddr}

\calclayout

\begin{document}

\title{Early onset of structural inequality in the formation of collaborative knowledge, Wikipedia}

\author{Jinhyuk Yun}
\address{Future Technology Analysis Center, Korea Institute of Science and Technology Information, Seoul 02456, Korea}

\author{Sang Hoon Lee}
\address{Department of Liberal Arts, Gyeongnam National University of Science and Technology, Jinju 52725, Korea}
\email[Corresponding author]{lshlj82@gntech.ac.kr}

\author{Hawoong Jeong}
\address{Department of Physics, Korea Advanced Institute of Science and Technology, Daejeon 34141, Korea; Institute for the BioCentury, Korea Advanced Institute of Science and Technology, Daejeon 34141, Korea}
\email[Corresponding author]{hjeong@kaist.edu}


\date{This manuscript was compiled on \today}

\begin{abstract}
We perform an in-depth analysis on the inequality in $863$ Wikimedia projects. We take the complete editing history of $267\,304\,095$ Wikimedia items until 2016, which not only covers every language edition of Wikipedia, but also embraces the complete versions of Wiktionary, Wikisource, Wikivoyage, etc. Our findings of common growth pattern described by the interrelations between four characteristic growth yardsticks suggest a universal law of communal data formation. In this encyclopaedic data set, we observe the interplay between the number of edits and the degree of inequality. In particular, the rapid increase in the Gini coefficient suggests that this entrenched inequality stems from the nature of such open-editing communal data sets, namely the abiogenesis of the supereditors' oligopoly. We show that these supereditor groups were created at the early stages of these open-editing media and are still active. Furthermore, our model considers both short-term and long-term memories to successfully elucidate the underlying mechanism of the establishment of oligarchy in Wikipedia. Our results anticipate a noticeable prospect of such communal databases in the future: the disparity will not be resolved spontaneously.
\end{abstract}

\maketitle
\pagestyle{plain}

The world we are living in today is a result of an enormous amount of cumulated human knowledge. It is therefore essential to understand the process of knowledge creation, and, perhaps more importantly, the collaborative human behaviours behind it, to maintain and enhance the development of our society. Contrary to this, the quantified data for analysing human history have been mostly far from satisfactory, obviously because of the lack of systematically preserved records describing the details of human knowledge development. As a result, investigation of human knowledge creation and collaboration has long been anecdotal and it was the job of the historians and anthropologists to fill the gaps based on fragmented evidences found all over the place. However, the situation completely changed at the turn of the century. The emergence of information technology in this century has offered environments to share up-to-date information generated by everyone online. The segregating lines between information producers and consumers have blurred in modern society, which is called \emph{produsage}~\cite{Bruns2008}. People believe that such a new environment will bring in ``democratisation'' of knowledge~\cite{Lemke2009}. At the same time, this accumulation of an inconceivable volume of information produced by everyone who is online every second has generated an unprecedented scale of exhaustive records of digital footprints harbouring human knowledge creation processes. Wikipedia, a representative open editing knowledge, may be referred to as the department store for contributors who generate information~\cite{Wikipedia}. Of course, due to the nature of information online, there has been a perpetual doubt regarding its credibility, i.e., it is sometimes considered unstable, imprecise, and even misleading. On the other hand, studies have proved that the current state of accuracy of Wikipedia is remarkable; the accuracy of its contents surpasses that of traditional encyclopaedias~\cite{Chesney2006, Giles2005}. Nevertheless, researchers found significant heterogeneities in the editing processes, based on the monopoly or oligopoly of a few supereditors' who govern the contents; thus, it is still far from being the Elysium of communal knowledge that we desire it to be~\cite{Gandica2015, Yun2016, Heaberlin2016, Zha2016, Ortega2008}. 

However, the majority of such studies, including our own, focused on a few language editions, mostly the English edition of Wikipedia to examine the dynamics and the properties of the communal data set (an editable data set shared within a community to build collective knowledge)~\cite{Chesney2006,Yun2016,Kittur2008,Adler2008,Yasseri2012}. Although they successfully warned of the potential risks behind the current social structure in English editions, it is clear that cultural background affects the behaviours of individuals. For instance, people belonging to different cultural backgrounds tend to use different symbols in their Web usage~\cite{Barber1998}, and the design of Web pages is also affected by their backgrounds~\cite{Marcus2000, Schmid-Isler2000}. Similarly, the users of Wikipedia must be affected by their social norms or cultures. It is also reported that editors in different language editions edit Wikipedia in distinctive patterns~\cite{Pfeil2006}. Moreover, the linguistic complexity of English Wikipedia differs from those of the German and Spanish editions~\cite{SKim2016}. Therefore, the results based on cultural differences appear to deny the generality in establishing such heterogeneity. On the other hand, those studies are based on small samples obtained from non-identically sized data sets, e.g., the number of articles in the English Wikipedia and the number of articles in Wikipedia in other languages are of different orders of magnitude; the English Wikipedia has at least five times more edits than others. Thus, the result might be caused by their relative sizes, and it is impossible to accept or to deny the lack of generality. As a result, the discussion is limited to a vague regional generality among the English users; however, the panhuman-scale behaviour remains unexplained.

To investigate the aforementioned generality, we extend our previous analysis on the English Wikipedia~\cite{Yun2016} to all $863$  Wikimedia projects~\cite{Wikimedia}, which are composed of various types of communal data sets served by the non-profit organisation called Wikimedia Foundation to encourage worldwide knowledge collection and sharing. For this purpose, we investigate the heterogeneity in contribution and supereditors' share in the Wikimedia projects to understand the socio-psychology behind open-editing communal data sets. In particular, we inspect the complete editing history of every Wikimedia project to assess Wikipedia's growth. We mainly focus on the number of edits, the number of editors, the number of articles, the article sizes, and the level of heterogeneity displayed in the inequality index. We demonstrate that there exist typical growth patterns of such open-editing communal data sets that eventually establish drastic heterogeneity among the editors' contribution. In addition, to comprehend the mechanism behind such disparity, we introduce an agent-based model that replicates the interactions between communal data sets and editors. Our model takes into account the competition between the editors' natural decrement in motivation over time, the editors' stronger memory on more recent activities, and the psychological attachment to their articles. The model reproduces the actual universal growth patterns, which are consistent with real data.

\section*{Results}

\subsection*{Data Set}\label{sec:data_set}

For our analysis, we use the March 2016 dump of the entire Wikimedia projects~\cite{WikimediaDownloads}. This data set not only includes the well-known Wikipedia, but also covers its sibling projects such as Wiktionary, Wikibooks, Wikiquote, Wikisource, Wikinews, Wikiversity, Wikivoyage, etc., in different languages (see Supplementary Table~\ref{table01} for its detailed composition). Each of these open-editing projects has a distinct subject and object. For example, each language edition of Wiktionary aims to describe all words from all the other languages listed in the main language of the edition, e.g., the English edition of Wikitionary has descriptions of the words of all languages in English. The differences between the objects may yield the gaps in the editing behaviours of editors belonging to each project, caused by the difference in demographic pools, accessibility, degrees of interests, etc. This dump contains the complete copy of Wikimedia articles from the beginning of 15 January 2001 to 5 March 2016, including the raw text and metadata sources in the Extensible Markup Language (XML) format. 

In this data set, there are a total of $267\,304\,095$ articles across all Wikimedia projects with their complete edit histories. Each article documents either the Wikipedia account identification (ID) or the Internet protocol (IP) address of the editor for each edit, the article size, the timestamp for each edit, etc. A single character takes one byte, except for a few cases such as Korean (two bytes) and Chinese (two or three bytes), so the article size in bytes is a direct measure of the article length~\cite{Utf8}. Each data set contains a number of articles ranging from $43\,124\,816$ (Wikimedia Commons: a database of freely usable audiovisual media) to $3$ (Wikipedia Login: a database used for administrative purposes), the number of editors range from $44\,349\,908$ (English Wikipedia) to $5$ (Nostalgia Wikipedia: read-only copy of the old English Wikipedia), the number of edits range from $654\,163\,757$ (English Wikipedia) to $14$ (Wikipedia Login), and the total article size ranges from $99\,519\,138\,751$ bytes (English Wikipedia) to $1\,206$ bytes (Wikipedia Login). See Supplementary Fig.~\ref{fig:wiki_statistics} for the distributions of various characteristic measures. 

Previous studies, including our own, used a few language editions, commonly restricted to the English edition of Wikipedia, which is the largest~\cite{Chesney2006,Yun2016,Kittur2008,Adler2008,Yasseri2012}. In addition, Wikimedia projects other than Wikipedia are not usually analysed, even though several language editions of Wikipedia were considered~\cite{SKim2016, Hale2014}. It is true that most other Wikimedia projects are not as large as the English Wikipedia, as shown by the fat-tailed distributions for different Wikimedia projects in Supplementary Fig.~\ref{fig:wiki_statistics}. However, the properties of everyday phenomena vary by their sizes~\cite{WSong2003}; thus, the characteristics of such communal databases may vary by their sizes and categories. Therefore, these smaller editions should not be neglected because of their being much smaller than the English editions, if we aim to comprehend the omnidirectional properties of communal data sets. Accordingly, we consider all 863 editions of the Wikimedia projects for our analysis. 

Our main goal is to find the underlying principle in the development of communal data sets. The Wikimedia project, as the representative player among such communal data sets, consists of various types of data sets operated by the Wikimedia Foundation. This massive record of knowledge spans $273$ different languages and $12$ different types of subjects (see Supplementary Table~\ref{table01} for details). This variety allows us to explore the innate nature of human behaviour based on each of their written languages and purposes of use. We consider a single Wikimedia project as a sample of such communal data sets. In order to proceed with the in-depth analysis on the evolution of communal data sets, we stress the fact that most data sets are aged approximately $3.5 \times 10^8$ seconds (about eleven years; see Supplementary Fig.~\ref{fig:wiki_statistics_pdf}). Thus, most Wikimedia projects are of similar ages, suggesting that raw characteristic measures without time-rescaling are appropriate, as they are free from the age effect. 

\subsection*{Universality and Disparity for Communal Data Sets}\label{sec:universality}
In this section, we present the evidence of a universal growth pattern shared by all Wikimedia projects, as displayed by characteristic measures based on the current snapshot of the communal data set, such as the total number of edits $N_e$, the total number of editors $N_p$, the total number of articles $N_a$, and the current sizes $S$ (in bytes). Our primary interest is to identify the generality in growth of the communal data set, not individual articles. Thus, we use the total sum of the above values for all articles in a Wikimedia project, without considering the individualistic properties of its constituent articles. For example, $N_e$ is the total number of edits for a given edition of a Wikimedia project, or the sum of the number of edits of individual articles for that edition. Our first analysis on the interplay between such measures indicates their regularity regardless of age, language, and the type of data sets.

\subsubsection*{Growth Scale of Communal Data Sets}
\label{sec:growth_scale_of_communal_data sets}

\begin{figure*}
\includegraphics{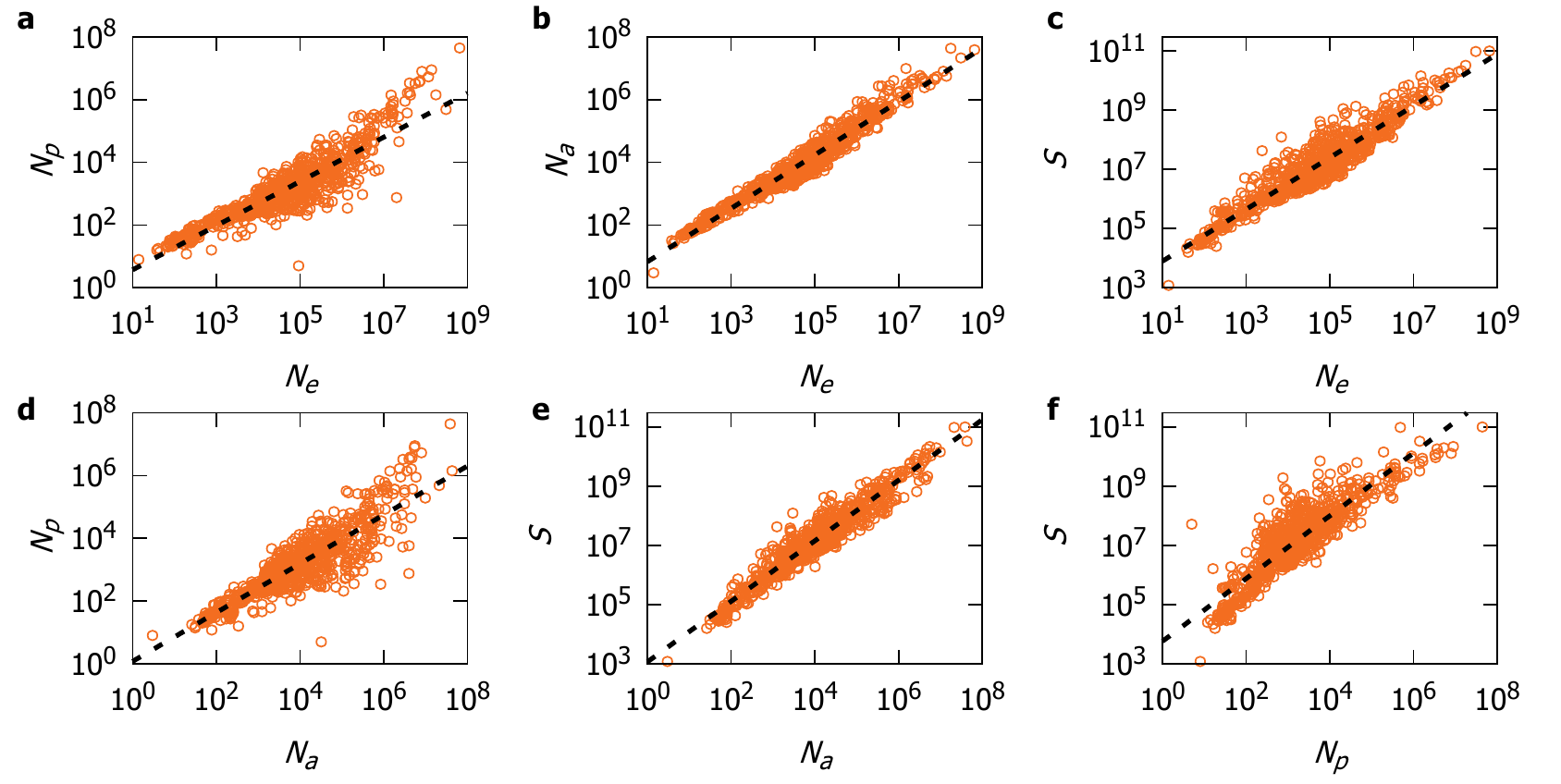}
\caption{The correlations between number of edits $N_e$, number of editors $N_p$, number of articles $N_a$, and total size of the data set $S$. Every correlation is characterized by the simple power-law growth form of $y \sim x^\lambda$. For the number of edits, the other measures grow sub-linearly with an exponent of \textbf{a}, $\lambda \simeq 0.70$ for the number of editors (Pearson correlation coefficient $\rho = 0.90$), \textbf{b}, $\lambda \simeq 0.85$ for the number of articles ($\rho = 0.85$), and \textbf{c}, $\lambda \simeq 0.87$ for the total size of data set in bytes ($\rho = 0.95$), respectively. \textbf{d}, The number of editors also sub-linearly increased by the number of articles with an exponent of $\lambda \simeq 0.78$ ($\rho = 0.65$), \textbf{e}, but the size of data set is almost linearly increased by the number of articles with an exponent of $\lambda \simeq 1.02$ ($\rho = 0.83$). Finally, panel \textbf{f}, displays the nearly linear interrelation ($\lambda \simeq 1.06$) between the number of editors and size of the data set, which in turn indicates that the average productivity of a single editor is maintained ($\rho = 0.73$). For the panels \textbf{a}--\textbf{f}, we estimate the power-law exponent with a simple linear regression methods in the double logarithmic scale (see Methods for the details). We present statistical details for the regressions in Supplementary Tables~\ref{table-nenp}--\ref{table-nps}.} 
\label{fig:wiki_statistics2}
\end{figure*}

We begin our analysis with the inspection of the inter-correlations between $N_e$, $N_a$, $N_p$, and $S$ in the current Wikimedia projects. One may speculate the absence of a general rule between measures due to the excessive heterogeneity of the current status (as shown in Supplementary Fig.~\ref{fig:wiki_statistics}) compared to its age distribution (as shown in Supplementary Fig.~\ref{fig:wiki_statistics_pdf}). As an example of the difference in different language editions of Wikipedia projects, it has been reported that the levels of language proficiency among editors in English Wikipedia are qualitatively different from those in the other language editions~\cite{SKim2016}. Despite such differences, we find common positive correlations between the measures. First, we observe a clear tendency that the number of editors, the number of articles, and the sizes of data sets gradually vary as the functions of the number of edits. The growing patterns are characterised by a simple sublinear growth of the form $y \sim x^\lambda$, where $x$ is the number of edits, and the exponent $\lambda \simeq 0.70$ (as shown in Fig.~\ref{fig:wiki_statistics2}\textbf{a}), $\lambda \simeq 0.85$ (as shown in Fig.~\ref{fig:wiki_statistics2}\textbf{b}), and $\lambda \simeq 0.87$ (as shown in Fig.~\ref{fig:wiki_statistics2}\textbf{c}), respectively. In other words, the frequency of appearance of new editors, that of brand new articles, and the increase in the amount of text slows down when more edits have taken place; from the perspective of editability, larger data set are more inefficient than smaller ones. 

To find the reason behind this stagnation in terms of the number of edits, we also track the interrelations between the other measures. The number of editors increases with the number of articles with the exponent $\lambda \simeq 0.78$ (as shown in Fig.~\ref{fig:wiki_statistics2}\textbf{d}). Meanwhile, the size of the article roughly linearly increases with the number of articles and the number of editors, with the exponents $\lambda \simeq 1.02$ and $\lambda \simeq 1.06$, respectively. In short, the rate of accumulation of the text remains almost constant regardless of the number of articles and the number of editors. One should note that our previous study on English Wikipedia suggests that i) the inter-event times between two consecutive edits in an individual article follows a universal distribution regardless of its age in real time and ii) the size difference between two consecutive edits also follows a universally right-skewed distribution regardless of the size~\cite{Yun2016}. Accompanied by the findings of our previous study, the result implies that the stagnation is caused by the decelerated appearance of new editors, and not the decreased productivity of the existing editors. In addition, in contrast to the interrelations between the measures we report here, the measures are not correlated with the age of the data sets (as shown in Supplementary Fig.~\ref{fig:wiki_statistics3}), indicating that the raw number of edits is a proper measure of time for comparison of various data sets rather than the actual time. As we observe in Supplementary Fig.~\ref{fig:wiki_statistics_pdf}, most of the Wikimedia projects are of similar ages; therefore, our analysis implies that the rate of growth per unit time also decreases as its size increases, as we have revealed in our previous study on English Wikipedia~\cite{Yun2016}. This universal growth scale is observed for all Wikimedia projects, regardless of their institutional aim, which implies that the common growth patterns is caused by the mutual nature of communal data sets (See Supplementary Fig.~\ref{fig:wiki_type}).

Along with the common scales observed in the different measures of a certain Wikimedia project, i.e., $N_e$, $N_a$, $N_p$, and $S$, it is also worthwhile to find possible scales between different types of Wikimedia projects in the same language. As we show in Supplementary Figs.~\ref{fig:wiki_type_ne_size}, \ref{fig:wiki_type_np_size}, \ref{fig:wiki_type_na_size}, and \ref{fig:wiki_type_S_size}, the bulks of different types of Wikimedia projects show almost identical tendency with respect to the size ratio. Each of the four measures for Wikipedia is considerably larger than that of any other types, and Wiktionary has the second largest size; other types of Wikimedia projects do not show great differences between them.

The next question is whether languages in Wikimedia projects can be categorised into distinct clusters that share growth patterns. To answer this question, we perform a cluster analysis by constructing a simple feature vector for a language, which is made up of characteristic measures from different types of Wikimedia projects (see Supplementary Methods for details). We present our results from two modern machine-learning techniques. In particular, we use the Dirichlet Process Gaussian Mixture Model~\cite{Blie2006}, which is known for its efficiency in partitioning the vectors in the case of unknown numbers of groups. Then, we use the t-SNE algorithm to visualise the higher dimensional feature vectors in the two-dimensional space, while preserving their original degrees of separations~\cite{Maaten2008}. As a result of the analysis, we could not observe any clear-cut clustering structures for different dimensions of feature vectors and different clustering parameters (Supplementary Figs.~\ref{fig:wiki_cluster_3vec}, \ref{fig:wiki_cluster_4vec}, and \ref{fig:wiki_cluster_5vec}). Therefore, our clustering results also support the existence of universal rules governing the growth of a communal data set, regardless of its language.

\subsubsection*{Wikimedia projects and Their Corresponding Socio-economic Indicators}\label{sec:wiki_socioeconomic}
We would like to check if there are any possible external factors that affect the current status of different Wikimedia projects in terms of the volume indicated by $N_e$, $N_a$, $N_p$, and $S$. The volume of Wikimedia projects is not simply determined by the total number of speakers of a language. As an illustrative example, the Spanish Wikipedia is approximately ten times larger than the Hindi Wikipedia, despite the fact that both Spanish and Hindi have around half a billion speakers each~\cite{Ethnologue2017}. Considering the fact that the ages of both Wikipedia language editions are comparable (16 years for Spanish Wikipedia and 14 years for Hindi Wikipedia), the growth of Hindi Wikipedia has been much slower than the Spanish Wikipedia up to this point. To elucidate the reason behind such a big difference, we inspect the factors affecting the current statuses of Wikimedia projects. The simplest factor is the ratio between the number of people using the language as a native language and those as a secondary language (see the Supplementary Methods for details). One may presume that the people using a particular language as a secondary language have less impact on the formation of communal data set written in that specific language, compared to its native language users. However, we find that the volumes of the Wikimedia project are more correlated with the volume of the secondary language users than the total volume of language users or the volume of users who are native speakers of the language (see Supplementary Fig.~\ref{fig:languser}). Other linguistic properties may also influence the growth rate of Wikimedia projects, so we tried categorising Wikimedia projects according to their written scripts. Rather surprisingly, there is no notable differences between the scripts (see Supplementary Fig.~\ref{fig:wiki_script}) because a single character in each script takes a different size in bytes, as we mentioned earlier.
 
In addition to the elementary linguistic factors, we try to consider more convoluted factors by cross-correlating the language editions of Wikimedia projects with the socio-economic statuses of countries to which the Wikimedia projects belong. We assign the dominant country of a certain language edition according to the following criterion: i) a country using the language as a primary or official language and ii) the first country in terms of the pageview share of the Wikipedia in that language (see Supplementary Methods for details). First, education levels of the countries show positive correlations with the volume of the corresponding Wikimedia projects, but not in a statistically significant manner (see Supplementary Fig.~\ref{fig:education}). In addition, the total population does not impact much on the statuses of the Wikimedia projects, whereas the total population of the Internet users shows a strong positive correlation with the volume of the Wikimedia projects (see Supplementary Fig.~\ref{fig:economy}). The Gross Domestic Product (GDP) is also well correlated with the growth of the Wikimedia projects (see Supplementary Fig.~\ref{fig:economy}). On the other hand, the GDP per capita is not correlated with the volume of the Wikimedia projects (see Supplementary Fig.~\ref{fig:economy}). In summary, the scale of the economy, which is partly reflected in the number of Internet users, affects the growth of Wikipedia projects.

Apart from the scale of the overall economy, we also observe that the national expenditures and products for research and development (R\&D) show a significant correlation with the current volume of the Wikimedia projects. Larger investors of research also tend to have larger Wikimedia projects compared to their smaller counterparts (see Supplementary Fig.~\ref{fig:rndinvest}). Consequently, the number of patents and the number of academic papers are also strongly correlated with the sizes of the Wikimedia projects (See Supplementary Figs.~\ref{fig:patent} and \ref{fig:paper}, and Supplementary Methods for the details of the patent and academic paper data set). Such an R\&D scale is determined by the national economic scale. Together with the results given above, the sizes of the Wikimedia projects are closely connected with the economic scale of a country in terms of total economic size, yet the per capita levels do not impact much on the current sizes of the Wikimedia projects. In other words, Wikimedia projects of richer countries grow faster and larger. 

\subsubsection*{Disparity in Contributions}
\label{sec:gini_editnumber}

The general growth patterns of the characteristic measures, $N_e$, $N_a$, $N_p$, and $S$, trigger an interesting proposition: could there also be a universal rule in the formation of recently reported structural heterogeneity~\cite{Yun2016,Heaberlin2016}? To examine the validity of the proposition, we employ the Gini coefficient, which is a conventional measure for inequality~\cite{Gini1912}. In our analysis, the Gini coefficient quantifies \emph{how the number of edits is distributed among different editors} involved in a certain Wikimedia project of interest, i.e., who have edited an article in the project at least once. The Gini coefficient ranges from $0$ for the minimal heterogeneity (or the maximal homogeneity---when every editor contributes equally) to $1$ for the maximal heterogeneity (when only a single editor contributes everything). We consider the number of edits and the data size for individual editors as the variables of interest in the Gini coefficient; these variables are referred to as ``wealth'', unless specified otherwise (as the Gini coefficient is usually used to quantify the inequality in economic wealth).  

The trend of the Gini coefficient as an increasing function of $N_e$ displayed in Supplementary Figs.~\ref{fig:wiki_gini} connotes that the disparity is intensified as the communal data set grows. Larger values of $N_e$ induce intenser disparity not only for the number of edits performed by the editors (as shown in Supplementary Fig.~\ref{fig:wiki_gini}, but also for the total data changes (in bytes) made by the editors (as shown in Supplementary Fig.~\ref{fig:wiki_gini}). This increasing trend is still valid when we perform addition (as shown in Supplementary Fig.~\ref{fig:wiki_gini}) and subtraction separately (as shown in Supplementary Fig.~\ref{fig:wiki_gini}). In addition, because the age of the article does not severely affect the heterogeneity (as shown in Supplementary Fig.~\ref{fig:wiki_gini}), the observation of the Gini coefficient is consistent with our observation that the age of the article does not affect the current state of the communal data sets. We predict that the heterogeneity will become severer if a given data set is edited more frequently. There is no notable distinction between written scripts of Wikimedia projects (see Supplementary Fig.~\ref{fig:wiki_script_gini}) and the institutional objectives for different Wikimedia projects (see Supplementary Figs.~\ref{fig:wiki_type_gini} and \ref{fig:wiki_type_gini2}). To sum up, we observe that the universal pattern of heterogeneity increased with the number of edits based on the current snapshot of the communal data set.

\subsection*{Evidences for the Establishment of the Supereditors' Dominance}
\label{sec:editorscartel}

\begin{figure*}
\includegraphics{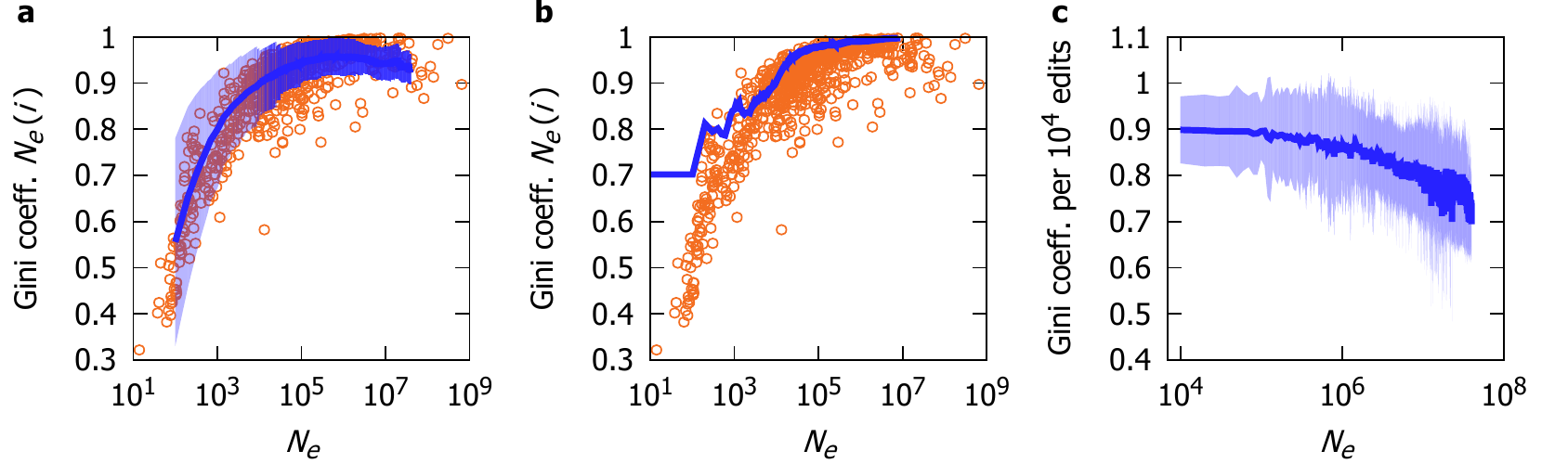}
\caption{The Gini coefficient of Wikimedia project as the functions of $N_e$. The blue curve in \textbf{a}, is the average Gini coefficient and the shaded area corresponds to its standard deviation, averaged over different Wikimedia projects. \textbf{b} shows a typical example of Cebuano Wikipedia. Initially, it does not seem to follow the general trend, the Gini coefficient for Cebuano Wikipedia starts to follow the trend curve for $N_e \gtrsim 10^4$. In panel \textbf{c}, we consider the number of edits for individual editors per unit time frame as the income variable in the Gini coefficient.}
\label{fig:wiki_gini2}
\end{figure*}

In the previous section, we have shown the current snapshots displaying the high level of heterogeneity and the increase in the Gini coefficient with the number of edits (see Supplementary Fig.~\ref{fig:wiki_gini}). Although the current statuses of all Wikimedia projects appear to follow a specific function of $N_e$, this snapshot could be coincidental. Thus, we further track the actual history to confirm or reject the hypothesis of possible coincidence so that we can judge if the increasing trend is actually the inherent nature of the formation of communal data sets. To check the hypothesis, we set the initial number of edits of all $863$ data sets to the same value [$N_e(t = 0) = 0$] and record the trajectories of the Gini coefficient as functions of $N_e$ (see Fig.~\ref{fig:wiki_gini2}\textbf{a} for the curve averaged over data sets with the deviation). Similar to the conventional usage of the Gini coefficient for wealth distributions, we use the cumulated number of edits up to $N_e$ (note that as discussed previously, the unit of time in this case is $N_e$) for each editor, which is a wealth variable. Note that, technically, the Gini coefficient is undefined when a single editor has edited a data set (as we define the set of editors as the editors who have contributed at least once), but we take the Gini coefficient as $1$ for that case because it well describes the completely monopolised state. Our result shows that the average Gini coefficient is coterminous with the current states of the Wikimedia projects (see Fig.~\ref{fig:wiki_gini2}\textbf{a}); thus the current status of a specific data set can be taken as a certain midpoint of a single master curve described as a function of $N_e$. For example, a history of the Cebuano Wikipedia clearly follows the typical growth pattern for $N_e > 10^4$ (see Fig.~\ref{fig:wiki_gini2}\textbf{b}), except for the initial fluctuations for small values of $N_e$.

Although we employ the Gini coefficient as the inequality measure in accumulated wealth distributions, an alternative approach of the index for an income is also widely accepted. In economics, the income is defined as the value gained within a specific time frame~\cite{Mankiw2014}. Alternatively, we also consider the number of edits for individual editors per unit time frame as the ``income'' variable in the Gini coefficient; this is called income unless otherwise specified. In other words, the wealth analysed before is the accumulated income from the onset of an individual editor's first activity. In this study, we use the time window of $10^4$ edits, but the different values of time frame do not affect the result meaningfully. The Gini coefficient in terms of the income distribution for the communal data set as a function of $N_e$ indicates that the larger $N_e$ values induce less severe heterogeneity in the contributions (see Fig.~\ref{fig:wiki_gini2}\textbf{c}). It indeed suggests that the income distribution becomes more homogeneous with time (see Fig.~\ref{fig:wiki_gini2}\textbf{c}), whereas the disparity in the wealth distribution is maintained (see Fig.~\ref{fig:wiki_gini2}\textbf{a}).

\begin{figure*}
\includegraphics{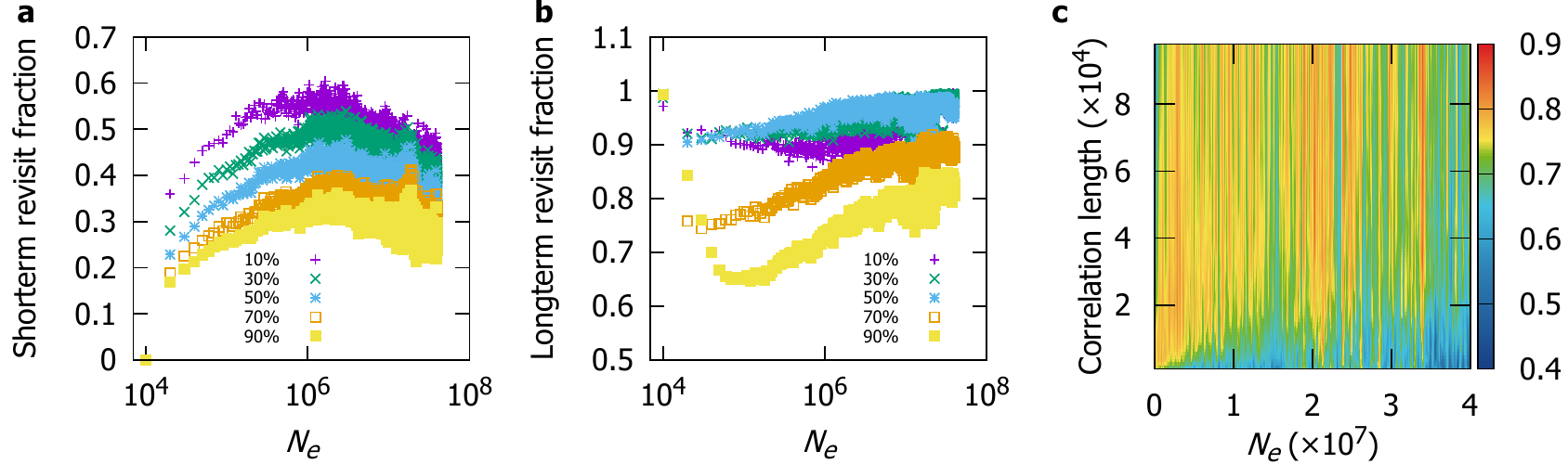}
\caption{The properties of the revisiting editors are characterized by their activity. The fraction of editors who are in a certain percentile within $\Delta N_e = 10^4$ edits, \textbf{a}, if the editors are also in a certain percentile for the previous $10^4$ edits and \textbf{b}, if the editors are also in a certain percentile for the entire edit activity until the specific point $N_e$. \textbf{c}, The Pearson correlation between the two lists of numbers of edits performed by a specific editor between previous $n$ edits and next $n$ edits for given number of edits on the horizontal axis, averaged over the Wikimedia projects. Here, $n$ is the value on vertical axis. We consider the number of edits of next (previous) $10^4$ edits for the editor as $0$ when the editor only appeared only in previous (next) $10^4$ edits, respectively. The correlation is undefined if only one editor is in both sequences, and we set the value as $1$ by convention in that case, because it corresponds to the complete dominance by that editor.}
\label{fig:wiki_revisit}
\end{figure*}

Therefore, the heterogeneity in the wealth distributions are intensified over time, whereas the disparity in the contributions per time frame become less severe with time for all editors. To consolidate the two results, we examine how the rich-get-richer concept affects the communal data set in details. Figures~\ref{fig:wiki_revisit}\textbf{a, b} suggest that the editors tend not only to keep their short-term social positions but also to maintain their long-term social positions. For instance, $58.1\%$ of the editors remain in the rank in the range of $[0\%,10\%]$ ($10\%$ from the top) for next $10^4$ edits if the editors were ranked in the range of $[0\%,10\%]$ ($10\%$ from the top) in the time window of $990\,000 \le N_e < 1\,000\,000$; meanwhile, only $32.6\%$ of the editors ranked in the range of $[0\%,90\%]$, i.e., except for the bottom $10\%$, retain their positions (see Fig.~\ref{fig:wiki_revisit}\textbf{a}). In other words, the editors who edit more often within a specific time window tend to edit more often later as well. Although the exact proportion and the number of edits for each percentile vary over time, the distinction between the social classes is preserved. As a result, a hierarchical structure between the editors is gradually becoming concrete. 

The trend is even clearer for the accumulated number of edits (see Fig.~\ref{fig:wiki_revisit}\textbf{b}). At the early stage, only highly ranked editors, whose amount of contributions are much larger than the median, maintain the positions represented by their cumulative number of edits up to that moment. Meanwhile, the rest of the lowly ranked editors, whose amount of contributions is much smaller than the median, change their positions more frequently. For every percentile, the percentile of revisiting editors gets more associated with the previous class over time, which eventually groups most editors under a stratified percentile. Therefore, rather solid classes are established at a very early stage and remain for a long time. The oligopoly of supereditors is thus visible~\cite{Yun2016, Heaberlin2016}. We have not only revisited the existence of such oligopoly, but also observed how its degree of influence changes as more edits are performed. The territories of these conglomerates extend beyond single articles and span the entire Wikimedia project level, and their leverage on Wikipedia is still growing (see Fig.~\ref{fig:wiki_revisit}\textbf{b}).

To comprehend the formation of such supereditors' massive share, we further discuss the interrelationship between the number of edits in two consecutive edit sequences in various time windows from the onset of the data set (as shown in Fig.~\ref{fig:wiki_revisit}\textbf{c}). We calculate the Pearson correlation coefficient between the lists of number of edits in two successive frame windows for an editor. Initially, two consecutive sequences of number of edits are highly correlated across various lengths of time windows, but the short-term correlation values are gradually decreased as more edits are performed. In addition, a boundary between the high-correlated domain (correlation $\gtrsim 0.7$) and low-correlated domain (correlation $\lesssim 0.7$) goes upwards with time; consequently, only long-term correlation is maintained. 

In the light of the above information, the results shown in Fig.~\ref{fig:wiki_revisit} explain the results shown in Figs.~\ref{fig:wiki_gini2}\textbf{a, c}; the disparity in the wealth distributions is preserved by the long-term correlations, whereas the disparity in the income distribution is steadily resolved due to the diminution of the short-term correlations. Although the short-term activities of editors may vary, the existence and dominance of a few editors are not resolved in the long run because of the dominance constructed at a very early stage of the communal data set. 

\subsection*{Other collaborative knowledge creations: the case of patents and academic papers}\label{sec:paper_patent}
One clear advantage of investigating online-based data such as Wikimedia projects is that we can identify individual contributions in the formation of the collective knowledge. On the other hand, the observation of ubiquitous growth scales and formation of strong heterogeneity also prompt an essential question: is early onset of heterogeneity caused by the inherent nature of online-based communal data set? In other words, is it also possible to find similar growth patterns in conventional knowledge formation processes? Although the Internet revolution leads online-based media to play an important role in constructing and spreading knowledge in this century, conventional platforms are a major route for disseminating expertise. To explore such a larger landscape of the collective knowledge formation, we extend our analysis to academic papers and patents, which are two pivotal media of traditional knowledge formations. 

For our analysis, we use patent data from the spring 2017 edition of European Patent Office (EPO) Worldwide Patent Statistical Database (PATSTAT) and academic paper data from 22 August 2017 dump of the entire SCOPUS CUSTOM XML DATA (see Supplementary Methods for details). For the patent data set, we assume that 91 distinct patent offices play roles analogous to the different editions of the Wikimedia projects. Similarly, we also considered each author's affiliated country as a unit of knowledge formation, analogous to an edition of the Wikimedia projects. Naturally, a single patent or a single scholarly article can be considered as an article in the Wikimedia projects. Unfortunately, it is impossible to trace the entire editing process during the composition of a single patent or a single article. Accordingly, we only use the information of the number of patents/articles and number of participants for each country. In addition, in contrast to the Wikimedia project, the time frame of our patent/paper data set does not cover the very beginning of the platform. Considering the long history of patents and academic papers, we could examine only a small contemporary subset of them, specifically, from 2000 for the patents and from 1996 for the papers. 

Based on our analysis of the Wikimedia projects, one may expect the existence of a general rule between the number of participants and the number of patents/articles. We find strong positive correlations between the measures for both patents and scholarly articles (Supplementary Figs.~\ref{fig:patent_gini} and \ref{fig:paper_gini}). Specifically, we find the Pearson correlation coefficient to be $\rho = 0.85$ between the number of patents for the number of inventors (who originally designed the technology), whereas it is $\rho = 0.74$ for the number of applicants (who originally filed the patent for the intellectual property rights). Statistics of academic papers show larger Pearson correlation coefficient $\rho = 0.97$ between the number of articles and the number of authors.

Our finding of general growth patterns across conventional knowledge platforms and online-based platforms encourage us to seek possible inequalities among participants of conventional knowledge formation platforms as well. We employ the Gini coefficient again to measure the degrees of inequality between the players in conventional knowledge media. Similar to the results of the Wikimedia projects, the heterogeneity levels of both patents and papers grow as increasing functions of the number of participants and number of research outputs. In contrast to the steep increment observed in the Wikimedia projects, patent and paper data sets show more gradual rising (compare Supplementary Figs.~\ref{fig:wiki_type_gini2}, \ref{fig:patent_gini}, and \ref{fig:paper_gini}), yet the Gini coefficient already reaches a high level ($\simeq 0.8$) for the leading countries in terms of the number of research output. In summary, an early rise in the disparity between the participants is a unique phenomenon of online-based communal data sets, whereas growing disparity seems to be a common nature of human knowledge formation. 

\subsection*{Agent-based Model of Heterogeneity Formation}\label{sec:memoryeffect}
To elucidate the dynamics in the formation of the supereditors' oligopoly, we introduce an agent-based model by importing different types of editors' ``memory'' affecting the motivation for edits. We assume that there are two fundamental and inherent motivations decaying over time, which govern the short-term and long-term behaviours of the editors. Our primary purpose is to examine the separate effects of these two memories governing the current state of Wikimedia projects. Besides these two decaying factors, the editors are also engaged in certain articles when they have already put in more effort in editing the articles~\cite{George2004}, which represents their psychological attachment to the articles. In the following, we describe in detail how we implement the socio-psychological effects into our mathematical model.

\subsubsection*{Model Description}\label{sec:model_description}
Our observation is mainly based on the indicators as functions of $N_e$ in the previous section, 
as we have already observed its validity with respect to the real data. Accordingly, we set a single edit as the unit of time $t$. The model begins with a single agent. Each agent represents a single editor who participates in the editing processes. We considered a single media representing the communal data set, or a single language edition of a certain Wikimedia project. In our model, we consider the action of editors to be motivated by their inherent natures, and introduce the parameters for the editors to describe their activities. First, for an editor $i$, we denote the accumulated number of edits as $N_i(t)$ at time $t$. The time of the first edit by the editor $t_{b;i}$ and time when the last edit occurred $t_{e;i}$ are specified. 
The dynamic rules are as follows. For each simulation step, the debut of a new agent and the revisit (or re-edit) by an already existing agent occur in turns. For every simulation step, a new agent appears with a constant probability $b$ and begins to participate in the editing process. Once a new agent appears in the data set at time $N_e$, the agent edits the data set at the time of inauguration so that $t_{b;i}$ and $t_{e;i}$ are assigned as $N_e$, and the time unit $t$ is increased by $1$ (the unit of the edit number). One should note that the time scales of $N_e$ for model and data are not identical, because model time scale can vary with the system size and parameters.

In the second step, an editor is chosen uniformly at random attempts to edit the data set. There are many factors affecting the motivation for the edits, but we take three: the long-term decay of motivations, the short-term motivation of ownership, and the psychological engagement of editors. In general, editors are highly motivated at the beginning of the participation, but their motivation fades steadily~\cite{Crane2008,FWu2007}. Thus, participants lose their attention as time goes by, which is modelled by the power-law decay as the factor $(t - t_{b;i})^{-k}$, where $k$ is the characteristic exponent representing the decay in motivation, which is observed in many temporally varying systems~\cite{Karsai2012,Karsai2014}. In addition, a fat-tailed distribution is observed for the time between the consecutive edits~\cite{Yun2016}, which suggests that the editing time scale of Wikipedia shows ``bursty'' behaviours, meaning that there is a short-term stimulation of edit motivation affected by the interval between an editor's latest editing $t_{e;i}$ and the current time $t$~\cite{Karsai2012, HJo2015}. This short-term stimulation of motivation is modelled as the factor $[1 + e^{-(t - t_{e;i})/\tau}]$, where $\tau$ is the characteristic time of this stimulation. Finally, there is a tendency for editors to be engaged when they have already participated more frequently~\cite{George2004,Yun2016}. The number of edits is assigned as $1$ at the time of first participation of the editor and increases by unity every time an agent participates in the edit process so that it is equivalent to the number of edits $N_i(t)$ up to the time point $t$. 

Taking these factors together, in our model, when an agent $i$ is chosen for editing, she participates in the editing with the probability
\begin{equation}
P_i [t; N_i(t), t_{b;i}, t_{e;i} ] = \min\left\{ 1, N_i (t) (t - t_{b;i})^{-k} [1 + e^{-(t- t_{e;i})/\tau}] \right\} \,.
\end{equation}

Once she decides to participate, $t_{e;i}$ is newly set as $t+1$ and $N_i(t+1) = N_i(t) + 1$. In addition, we also include the possibility for an agent to leave the editing process indefinitely. We consider that this departure is based on the loss of motivation to edit~\cite{Crane2008, FWu2007}. Therefore, in our model, an agent leaves the system when she chooses not to edit and $P_i [t; N_i(t), t_{b;i}, t_{e;i} ] < r$, where $r$ is a preassigned cut-off parameter common to all the editors. In the following section, we give some evidences that the formation of the current inequality is because of the above factors, regardless of the innate nature of an individual editor.

\subsubsection*{Model Results}\label{sec:model_results}

\begin{figure*}
\includegraphics{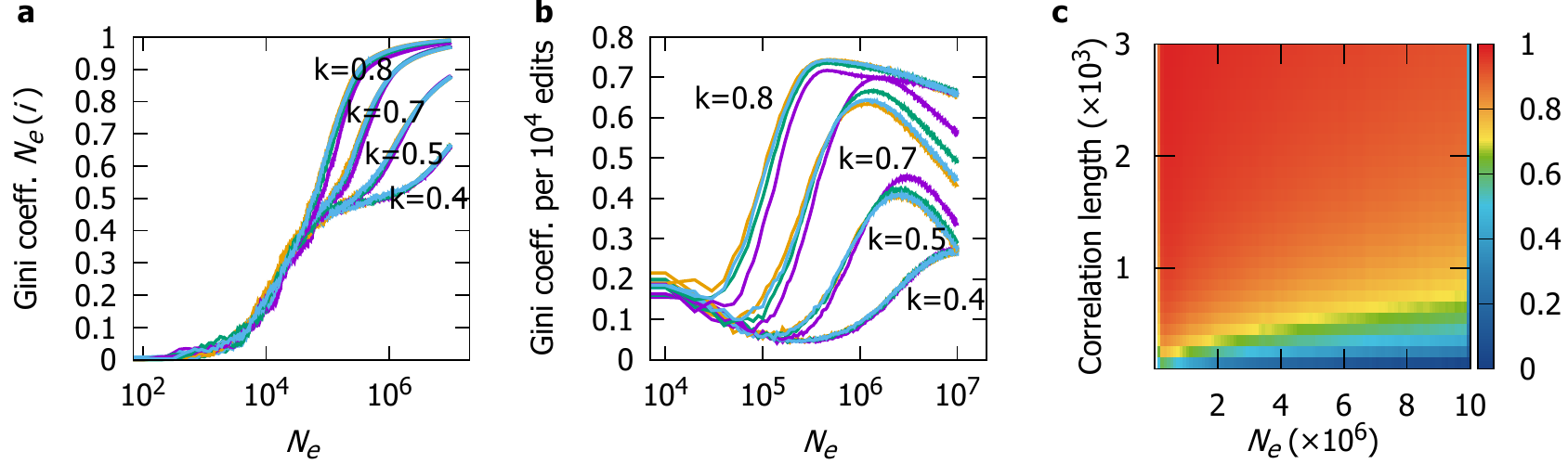}
\caption{The Gini coefficient from our model as functions of the number of edits. Panel \textbf{a} shows the Gini coefficient for the number of edits. Panel \textbf{b} shows the Gini coefficient for the time frame of every $10^4$ edits, for a given value in the horizontal edits. For panels \textbf{a, b}, the colour corresponds to the different value of $\tau$ ranges from $\infty$ (no short-term stimulation) to $0.001$: $\tau = 0.01$ (purple), $\tau = 0.001$ (green), $\tau = 0.0001$ (blue), and $\tau \to \infty$ (yellow). For panels \textbf{a, b}, we use the following parameters: $b = 0.0001$, and $r = 0.01$. (\textbf{c}) The Pearson correlation between the lists of numbers of edits performed by an editor between previous $n$ edits and next $n$ edits for the number of edits on horizontal axis, where $n$ is the value on the vertical axis. We consider the number of edits for next (previous) $10^4$ edits for an editor as $0$ when the editor only appears in the previous (next) $10^4$ edits, respectively. For the panel (\textbf{c}), we used the following parameters: $b = 0.0001$, $k = 0.8$, $\tau \to \infty$, and $r = 0.01$. We check that other choice of $\tau$ gives similar results, but we show the result with $\tau \to \infty$ to emphasise the long-term correlation. For (\textbf{a--c}), each parameter is averaged over 1000 independent realizations.}
\label{fig:wiki_model}
\end{figure*}

Previously, we have shown the increasing trend of the Gini coefficient as the number of edits is increased, which is, in particular, rapidly increased at the early stage of the data set and stabilised at the high level (the Gini coefficient $\gtrsim 0.8$ for $N_e \gtrsim 10^4$, see Fig.~\ref{fig:wiki_gini2}). Our model result is consistent with the empirical observations. The Gini coefficient of the model data set is rapidly increased until the high level is reached at $N_e \simeq 10^5$ for $k = 0.8$ (compare Fig~\ref{fig:wiki_model}\textbf{a} with Fig.~\ref{fig:wiki_gini2}\textbf{a}). Smaller $k$ values yield a slower increment in the Gini coefficient, whereas the $\tau$ value does not affect it significantly. The Gini coefficient does not reach the high level (the Gini coefficient $\simeq$ 1) if we assign $k \gtrsim 1$, which suggests that a moderate decaying of motivation is essential to reproduce the current state of the communal data sets. The Gini coefficient of the income distribution also displays results from our model that are similar to that from the data. For $k = 0.8$, the Gini coefficient for the income is steadily decreased from $N_e \simeq 10^5$ (see Fig~\ref{fig:wiki_model}\textbf{b}), which is observed in the data for $N_e \gtrsim 10^5$ (see Fig.~\ref{fig:wiki_gini2}\textbf{c}). 

One may claim that the early tendency of our model ($N_e \lesssim 10^5$) for the Gini coefficient for income disagrees with the data. In addition, one may observe undulation points, which is absent in the data, at $N_e \simeq 2\times10^4$ for the Gini coefficient for wealth in cases other than $k = 0.8$. Although our model reproduces the patterns in the data at the later stage ($N_e \gtrsim 10^5$ for $k = 0.8$), we would like to remark on this seemingly different growth pattern at the early stage. First, as a minimalistic model, we do not mean to reproduce the inherent disparity from the agents' attributes, e.g., social class, education level, and language fluency. Furthermore, we do not mean to explain the data all the way from the complicated early procedure when people launch a new project which is exposed exclusively for limited users, such as i) \textit{language proposal} and ii) \textit{incubator}. Naturally, the dynamics of this early stage is very different from the public launch. Despite such difficulty, our model starting with a ``regular'' dynamics from the very first agent also shows such an early stabilizing period indeed. The relation between the number of editors and the number of edits in the model also displays two different stages (see Supplementary Figs.~\ref{fig:model_npne_0_4}--~\ref{fig:model_npne_0_8}). Although the transition point between two stages varies by the specific values of $k$, we assume that this point may correspond to the undulation points for the Gini coefficients for wealth and income (compare Fig.~\ref{fig:wiki_model} with Supplementary Figs.~\ref{fig:model_npne_0_4}--~\ref{fig:model_npne_0_8}). To conclude, the apparent mismatch for the early stage between model and data is caused by the simplified early stabilizing stage for the model, which is difficult to capturing the very specific details in the real incubating stage for the Wikimedia projects.

In addition to the Gini coefficient, our model also reproduces the trend of reduced short-term correlations for the number of edits between time windows reported in Fig.~\ref{fig:wiki_revisit}\textbf{c}. As shown in Fig.~\ref{fig:wiki_model}\textbf{c}, the interrelationship between the number of edits in two consecutive sequences in various time frames from the onset of the data set gives a similar result. In the model and the real data, we observe a significant correlation between two consecutive sequences regardless of the lengths of the sequences. With time, the short-term correlation is steadily reduced, whereas the long-term correlation is sustained. Similar to the data, the border between large-correlation (correlation $\gtrsim 0.7$) and small-correlated domains (correlation $\lesssim 0.7$) rises as more edits are performed (see Fig.~\ref{fig:wiki_model}). The slope of a border is different for different $k$ values, but $\tau$ does not affect the slope. 

In short, the parameter $k$ mainly governs the overall dynamics despite the fact that the rapid increment in wealth inequality happens at the early stage and the gradual decrement in the income inequality always occurs. In other words, the loss of long-term motivation induces the inequality, whereas the short-term memory does not affect the system much. Therefore, the rich-gets-richer effect is mainly driven by the accumulated engagement induced by previous edits, and such a long-term engagement leads to the formation of the supereditors' oligopoly lasting to date. In addition, our model indicates that the supereditors' oligopoly can be formed without the direct communication between editors or, in other words, any direct pressure from the society. 

\section*{Discussion}\label{sec:conclusion}
In this study, we have examined the common patterns in the communal data sets displayed in all language editions of different types of Wikimedia projects. Although some studies have uncovered the general patterns before, it is usually based on partial observations of data sets of specific type or specific languages, which have left many unanswered questions and speculations~\cite{Chesney2006, Yun2016, Kittur2008, Adler2008, Yasseri2012}. However, the extensive data set formed from all Wikimedia projects, which record large-scale collaboration for the creation of collective knowledge, has given us an unprecedented opportunity to explore collaborative behaviours of human beings quantitatively. In this data set, we have observed the universal interplays between the number of editors, the number of articles, the number of edits, and the total length of the articles, which are characterised by the power-law scaling form with a single set of exponents. The existence of the universal growth rules in all the $273$ languages and $12$ types of Wikimedia projects suggests the panhuman-scale behaviour with regard to collaboration. 

This universal pattern is seen not only in the external appearances of the data sets, but also in their heterogeneity quantified by the Gini coefficient; the disparity is formed at a very early stage of the communal data sets and continues forward. It was widely hoped that the communal data sets will bring democratisation of knowledge~\cite{Lemke2009}, yet studies reveal that the current Wikimedia projects are hampered by strong heterogeneity in editing~\cite{Yun2016, Heaberlin2016}. We have demonstrated that the heterogeneity between the editors can be more deep-rooted than expected. The existence of the supereditors' massive share is a universal phenomenon across all communal data sets, i.e., Wikimedia projects, regardless of their sizes and activities. We have also observed the universal trend of intensified disparity for all types of data sets, which suggests that the vast share of a few dedicated editors will be intensified further. The value of such dedicated editors should be honoured because their voluntary dedication has indeed archived the current level of accuracy in Wikipedia~\cite{Chesney2006,Kittur2008a}. However, on the other hand, biased narrations on the topics were indeed reported and lost diversity may intensify the issues of systematic biases~\cite{Hube2017,Callahan2011,Reagle2011}, notwithstanding Wikipedia's continuing efforts toward neutral point of view. In addition, we have shown that a social stratum of such communal data sets can be formed at the very early stage and the polarisation of editors is already in process. 

Our study is not limited to the diagnosis of the current state of Wikimedia projects, but provides a general insight on the future direction of communal data sets. For instance, our simulation suggests that the inequality can be formulated without direct interactions between editors. Indeed, it is also observed that the editors tend to obey pre-established authorities~\cite{Heaberlin2016}. Again, undoubtedly, we acknowledge the dedication of such supereditors for maintaining the high-standard quality of the current state of Wikipedia by their (by definition) large amount of contribution to it~\cite{Chesney2006,Kittur2008a}. However, the total productivity of each editor is decreased as the number of edits is increased, which may result in less productivity and even less accuracy in the future. It was already reported that the growth of Wikipedia has slowed down~\cite{Suh2009}, and our analysis also warns that the inequality will not be easily resolved without active efforts. 

Since the turn of the century, Wikipedia has served as a spearhead of the international open knowledge market. However, strategic actions considering the nature of such a social structure are required to sustain the abundant playground for worldwide collaborations. Giving incentives to new editors may help this situation, but a suitable tutorial system that prevents vandalism and encourages decent editing activities is also needed. Fork and merge system commonly used in open-source communities also improve the editability of Wikimedia projects by serving as a secondary method of talk pages where editors can debate. With the Fork system, new editors fork their own versions to show their ideas, and it can be merged with the original article with debating.

Our finding displays abiogenesis imbalances in the formation of a particular set of communal data, but the result and implication of our study can be applied outside the Wikimedia projects. With the Internet penetrating our life deeply, online-based collaboration environments have become a mainstream platform. Therefore, interests on the contribution patterns in various communal datasets such as open-source and free software movements attract public attention~\cite{Gherardi2013, Tsay2014, Padhye2014, Benkler2006}. Compared to traditional (offline) collaboration system, the fruit of online collaboration system is immediately released as products in a collaborative fashion. Analysing millions of the outcomes in GitHub, Apache, GNU \& Free softwares, copyleft movements will display more detailed landscape of collaborative knowledge creations, which we will leave for future study. There will be extensive applications for understanding the collective behaviours of humankind based on this type of analysis, which, we hope, could give clues to solve social inequalities of even larger scale.

\section*{Methods}\label{sec:methods}

\subsection*{Data Description}
The data were obtained by downloading the March 2016 dump of all Wikimedia projects~\cite{WikimediaDownloads}. This data set include Wikipedia and its sibling projects such as Wiktionary, Wikibooks, Wikiquote, Wikisource, Wikinews, Wikiversity, Wikivoyage, etc., in different languages. Our data contain $267\,304\,095$ articles across all Wikimedia projects with the complete history of edits and the Wikipedia account identification (ID) or Internet protocol (IP) address of the editor for each edit, the article size, the timestamp for each edit, etc.  

\subsection*{Estimation of power exponent for the correlation between the measures $N_e$, $N_p$, $N_a$, and $S$}
To estimate the power-law scaling relations between the measures, we apply a simple linear regression method to the logarithm of the values of interest with the assumption of the simple power-law scaling $y = Cx^ \lambda$. Inevitably, there are various types of noises and fluctuations in empirical observations, so the distribution should be in fact written as $ y = k(x/x_{\min})^{\lambda} + \kappa + \eta$, where $x_{\min}$ is the minimum value of $x$ from which the power law is observed, $\kappa$ is the constant background offset, and $\eta$ is the term for random fluctuations. We neglect those impacts to obtain overall collective tendency for the entire Wikimedia projects. This simple method has clear advantages over the complex multivariate regression: it is less sensitive to the heterogeneous disparity in empirical distribution. The aforementioned power-law scaling can be transformed as $\ln y = \lambda \ln x + \ln C$, and we perform the simple linear regression on these logarithmic values $\ln x$ and $\ln y$ to yield the exponent $\lambda$ and the proportionality constant $C$ (see Supplementary Tables~\ref{table-nenp}--\ref{table-nps} for the statistical details).

\subsection*{Model Construction}
Our model begins with a single agent accompanied with a single media representing the communal data set. For each simulation step, the entry of a new agent occurs (with the probability $b$). Then, an editor is chosen uniformly at random attempts to edit the data set. We consider three factors to determine the probability of editing. First, the motivation for performing the edit decays slowly by the power-law decay as the factor $(t - t_{b;i})^{-k}$, where $k$ is the characteristic exponent representing the motivation decay ~\cite{Karsai2012,Karsai2014}.  In addition, we include the short-term stimulation of motivation, which is modelled as the factor $[1 + e^{-(t - t_{e;i})/\tau}]$, due to consideration of ``bursty'' behaviours ~\cite{Yun2016}. Finally, the attachment to the edited articles increased by unity from an initial value of one every time an agent participated in the editing process, so that it is equivalent to the number of edits $N_i(t)$ at the time point $t$. 

Considering these factors together in our model, when an agent $i$ is chosen for editing, she participates in the editing with the probability
\begin{equation}
P_i [t; N_i(t), t_{b;i}, t_{e;i} ] = \min\left\{ 1, N_i (t) (t - t_{b;i})^{-k} [1 + e^{-(t- t_{e;i})/\tau}] \right\} \,. 
\end{equation}

\section*{Acknowledgements}
This work received institutional supports by Korea Institute of Science and Technology Information, and was supported by Gyeongnam National University of Science and Technology Grant in 2018--2019 (S.H.L.). The National Research Foundation (NRF) of Korea Grant funded by the Korean Government also supported this work through Grant NRF-2017R1E1A1A03070975 (J.Y.) and No. NRF-2017R1A2B3006930 (H.J.). The funders had no role in study design, data collection and analysis, decision to publish, or preparation of the manuscript.
 
\section*{Contributions}
All authors designed the experiment and wrote the manuscript. J.Y. collected and analysed the data.

\section*{Competing interests}
The authors declare no competing interests.

\section*{Corresponding author}
Correspondence to Sang Hoon Lee and Hawoong Jeong.

\pagebreak
\begin{center}
{\LARGE Supplementary Information}

\bigskip

Early onset of structural inequality in the formation of collaborative knowledge, Wikipedia

\end{center}
\setcounter{equation}{0}
\setcounter{figure}{0}
\setcounter{table}{0}
\setcounter{page}{1}

\makeatletter
\renewcommand{\figurename}{Supplementary Figure}
\renewcommand{\tablename}{Supplementary Table}

\section{supplementary Figures}\label{sec:sup_figs}
\null
\begin{figure*}[!ht]
\includegraphics{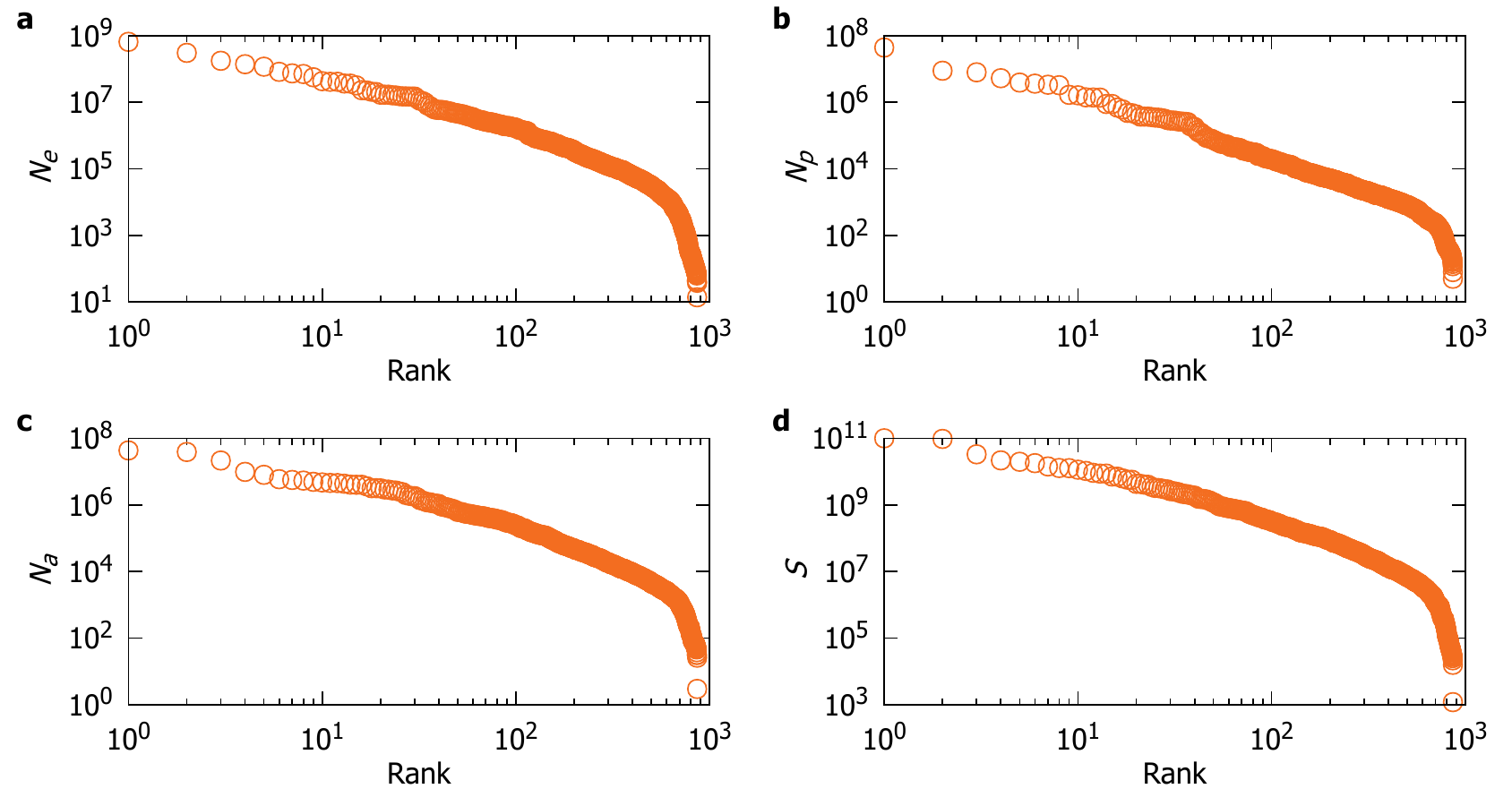}
\caption{The rank versus characteristic measures of Wikimedia projects: \textbf{a}, the number of edits $N_e$, \textbf{b}, the number of editors $N_p$, \textbf{c}, the number of articles $N_a$, and \textbf{d}, the total volume of texts $S$ (in the unit of bytes).}
\label{fig:wiki_statistics}
\end{figure*}

\pagebreak
\null
\begin{figure}[!ht]
\includegraphics{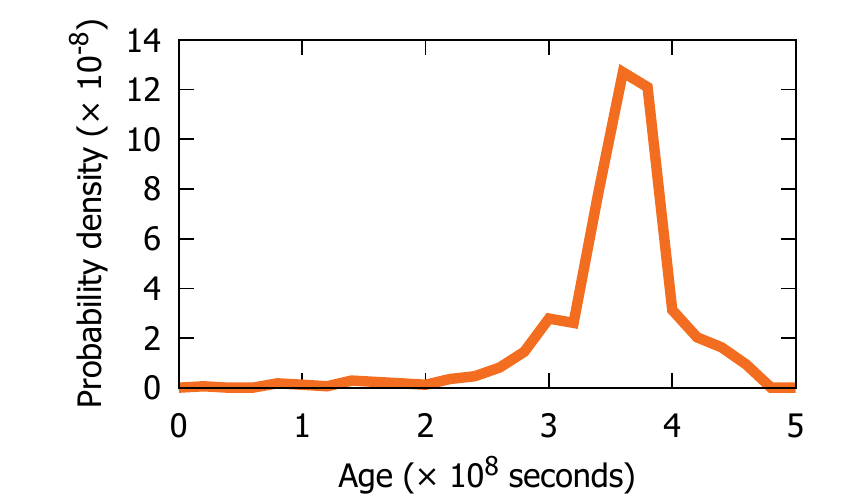}
\caption{The age distribution of Wikimedia projects, where we bin the data in the uniform length of $2\times10^7$ seconds (the resolution of the horizontal axis).}
\label{fig:wiki_statistics_pdf}
\end{figure}

\pagebreak
\null
\begin{figure*}[!ht]
\includegraphics{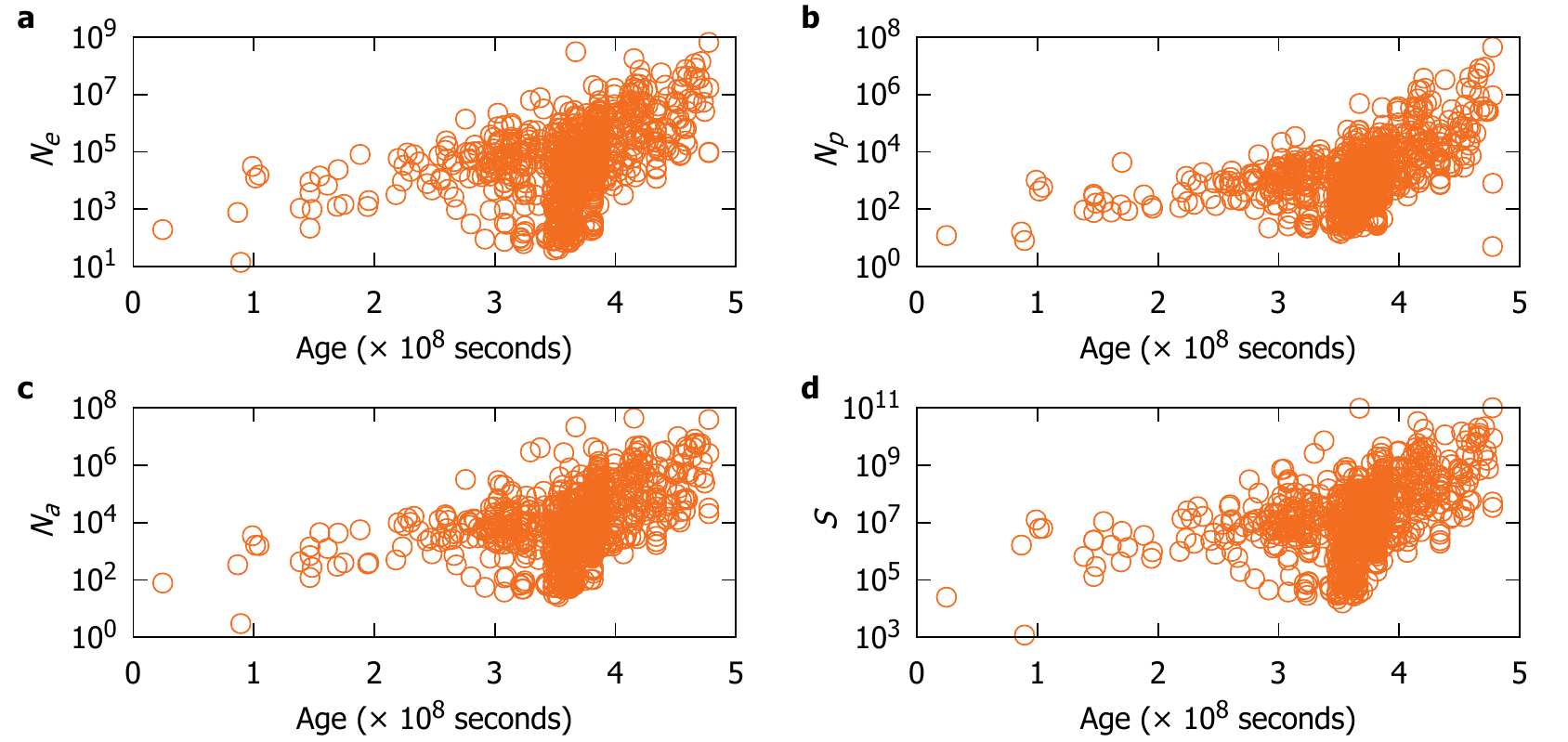}
\caption{The correlations for, \textbf{a}, the number of edits (Pearson correlation coefficient $\rho = 0.12$), \textbf{b}, the number of editors (Pearson correlation coefficient $\rho = 0.11$), \textbf{c}, the number of articles (Pearson correlation coefficient $\rho = 0.13$), and \textbf{d}, the total size of Wikimedia projects (Pearson correlation coefficient $\rho = 0.12$), with the age of Wikimedia projects.}
\label{fig:wiki_statistics3}
\end{figure*}

\pagebreak
\null
\begin{figure*}[!ht]
\includegraphics[width=\textwidth]{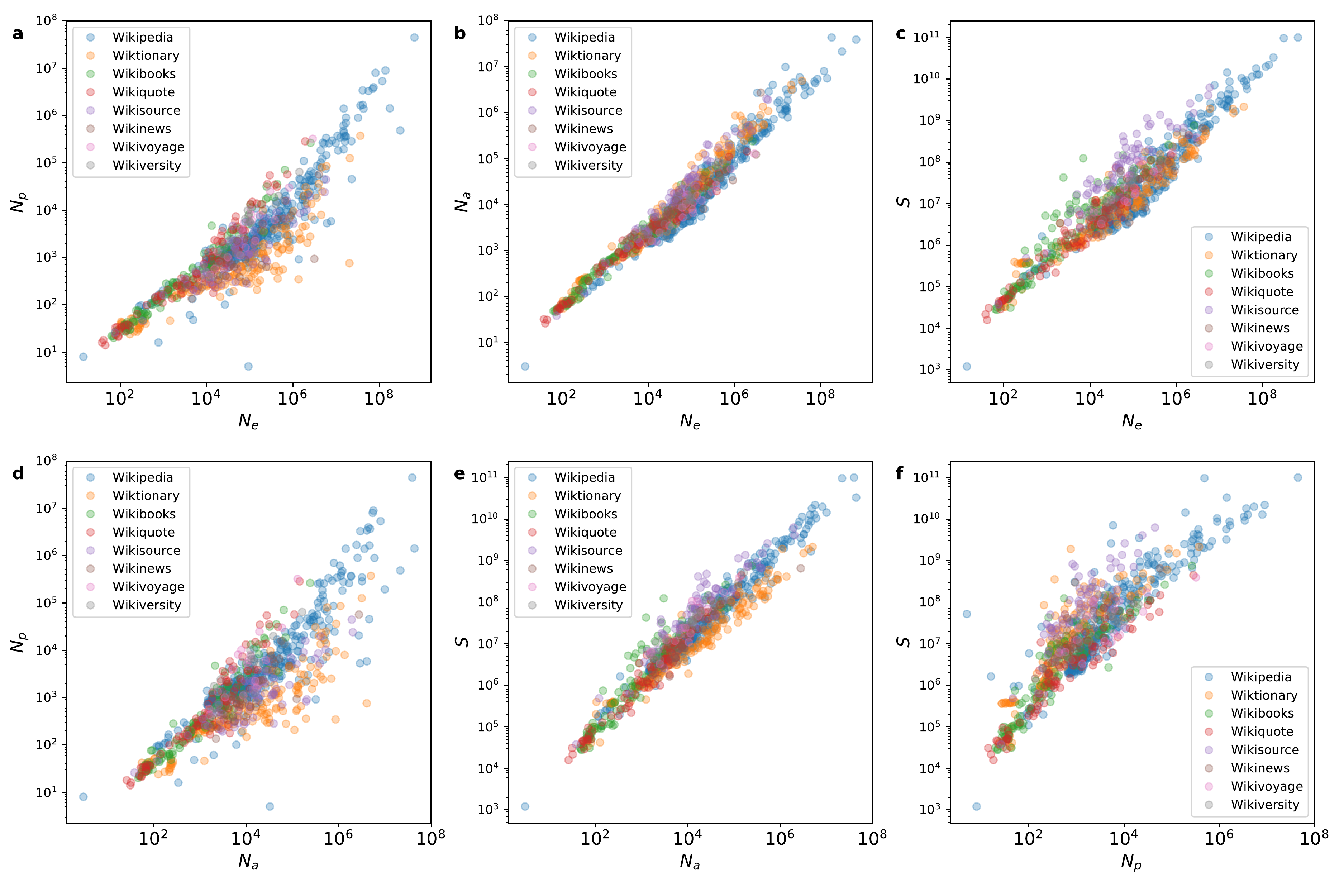}
\caption{The correlations between the number of edits $N_e$, the number of editors $N_p$, the number of articles $N_a$, and the total size of the data set $S$. The colours represent the types of Wikimedia projects.}
\label{fig:wiki_type}
\end{figure*}

\pagebreak
\null
\begin{figure*}[!ht]
\includegraphics[width=\textwidth]{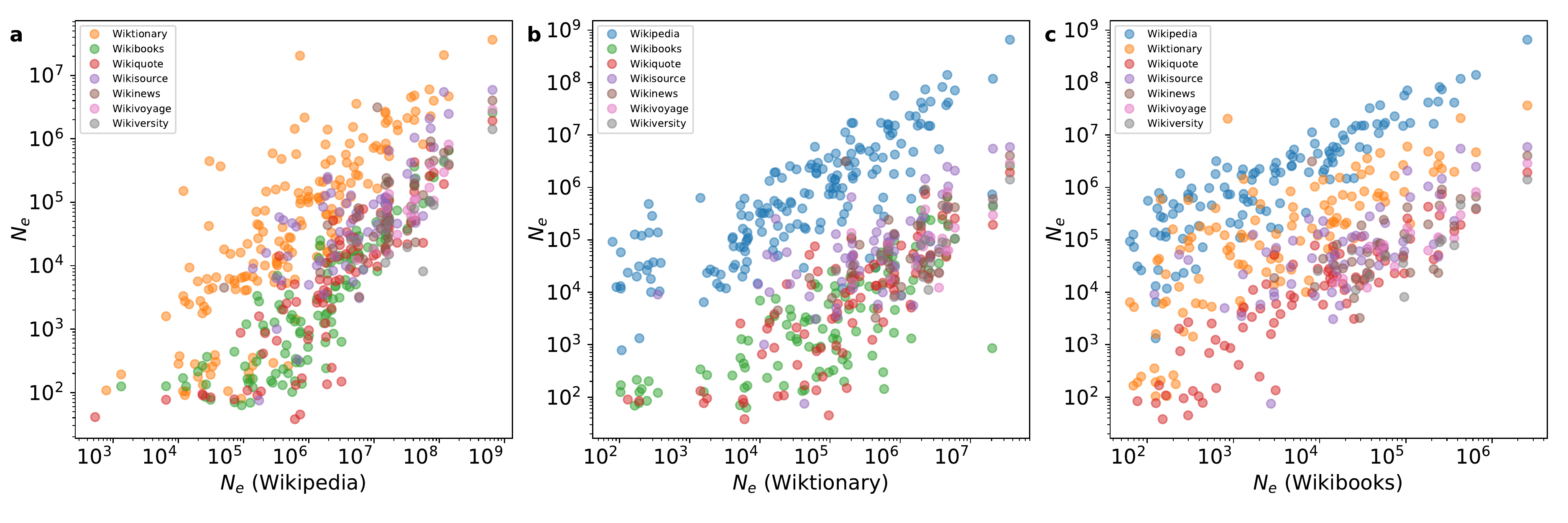}
\caption{The correlations between the number of edits $N_e$ of each Wikimedia projects and that of the three largest Wikimedia projects for each language. \textbf{a}, The correlation between Wikipedia and the others. \textbf{b}, The correlation between Wiktionary and the others. \textbf{c}, The correlation between Wikibooks and the others. The colours distinguish the types of corresponding Wikimedia projects. 
}
\label{fig:wiki_type_ne_size}
\end{figure*}

\pagebreak
\null
\begin{figure*}[!ht]
\includegraphics[width=\textwidth]{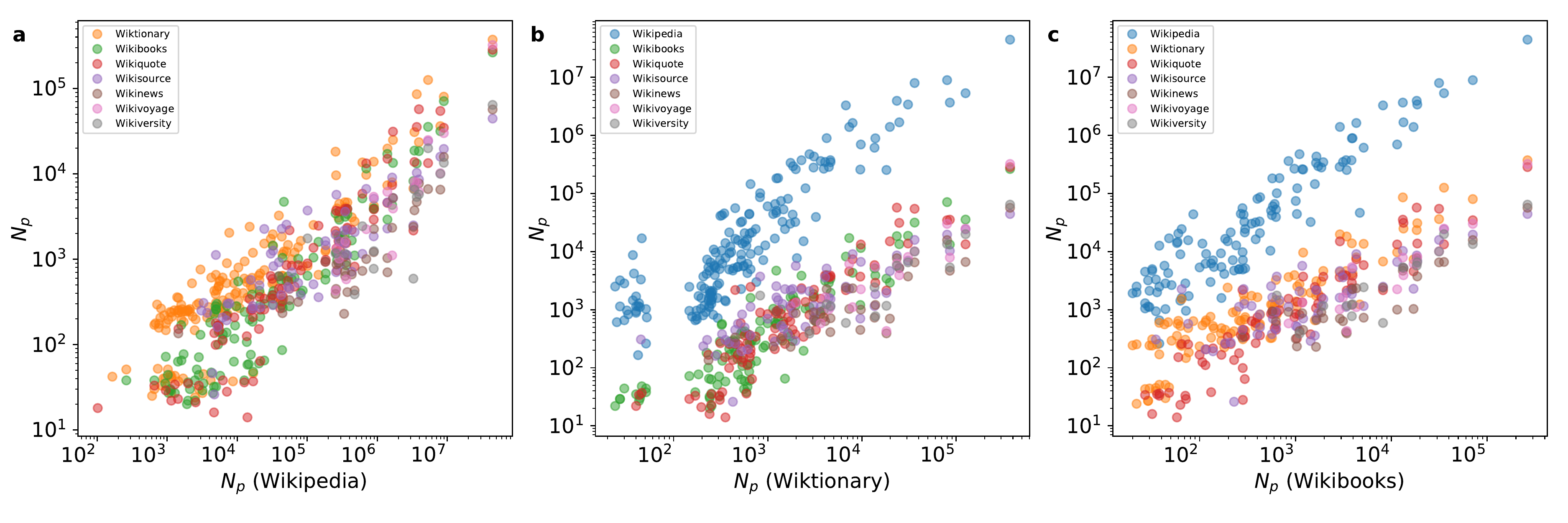}
\caption{The correlations between the number of editors $N_p$ of each Wikimedia projects and that of the three largest Wikimedia project for each language. \textbf{a}, The correlation between Wikipedia and the others. \textbf{b}, The correlation between Wiktionary and the others. \textbf{c}, The correlation between Wikibooks and the others. The colours distinguish the types of corresponding Wikimedia projects.
}
\label{fig:wiki_type_np_size}
\end{figure*}

\pagebreak
\null
\begin{figure*}[!ht]
\includegraphics[width=\textwidth]{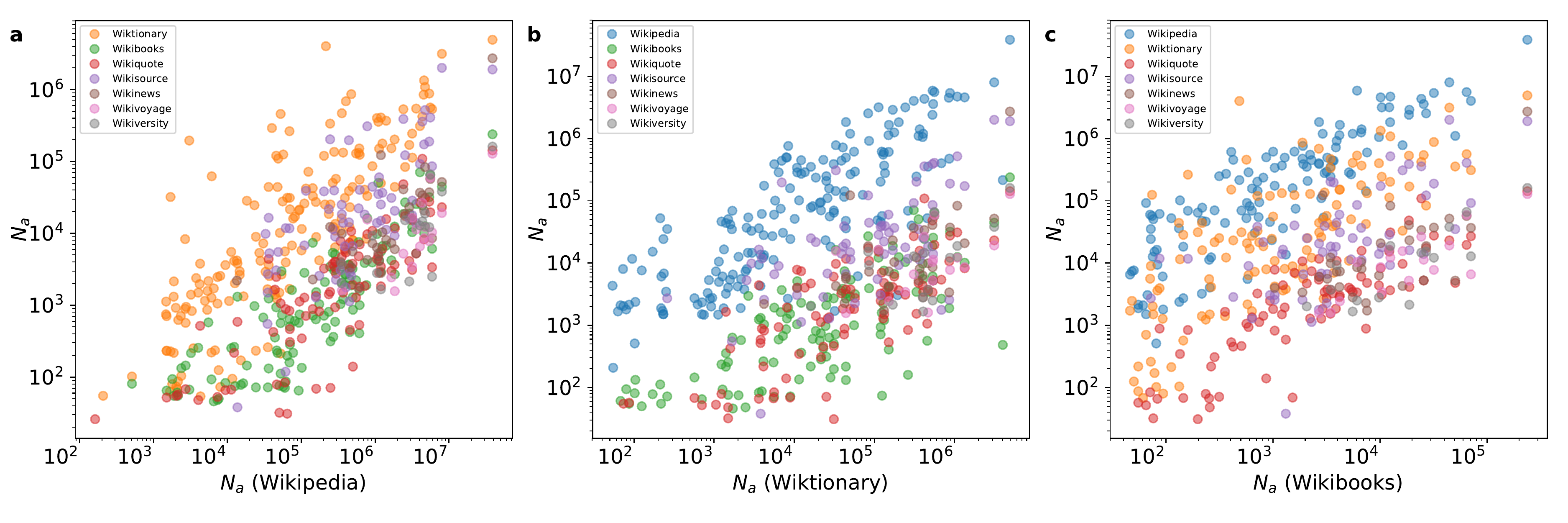}
\caption{The correlations between the number of articles $N_a$ of each Wikimedia projects and that of the three largest Wikimedia project for each language. \textbf{a}, The correlation between Wikipedia and the others. \textbf{b}, The correlation between Wiktionary and the others. \textbf{c}, The correlation between Wikibooks and the others. The colours distinguish the types of corresponding Wikimedia projects. 
}
\label{fig:wiki_type_na_size}
\end{figure*}

\pagebreak
\null
\begin{figure*}[!ht]
\includegraphics[width=\textwidth]{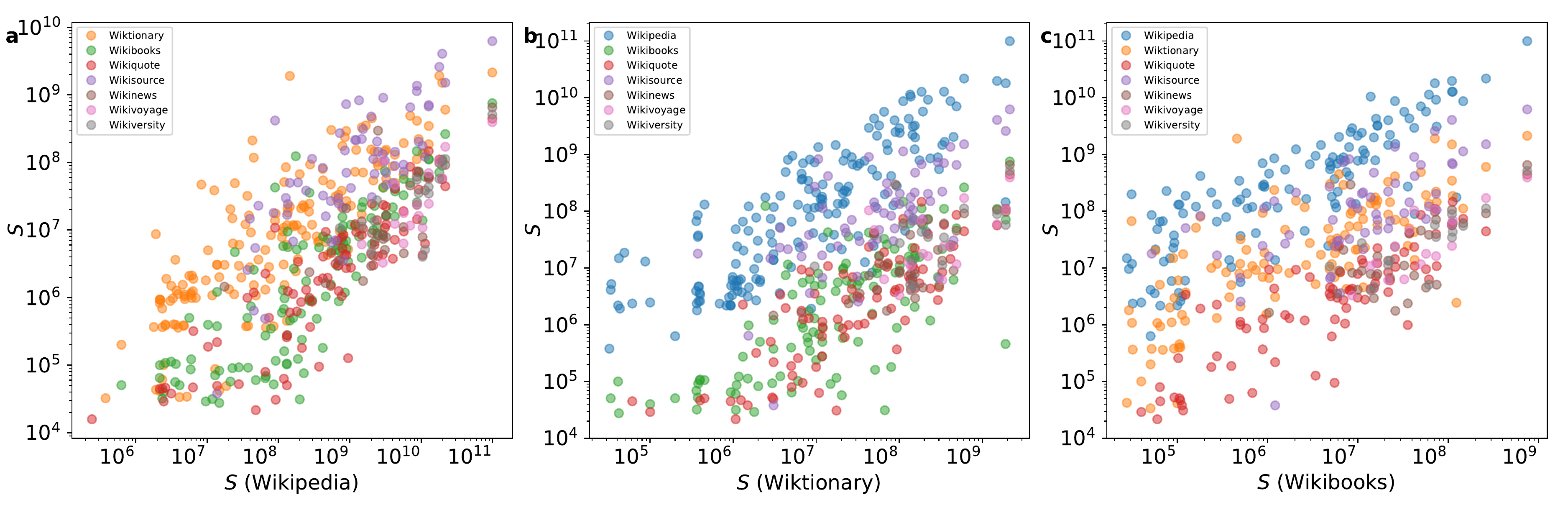}
\caption{The correlations between the total size of the data set $S$ of each Wikimedia projects and that of the three largest Wikimedia project for each language. \textbf{a}, The correlation between Wikipedia and the others. \textbf{b}, The correlation between Wiktionary and the others. \textbf{c}, The correlation between Wikibooks and the others. The colours distinguish the types of corresponding Wikimedia projects.}
\label{fig:wiki_type_S_size}
\end{figure*}

\pagebreak
\null
\begin{figure*}[!ht]
\includegraphics[width=\textwidth]{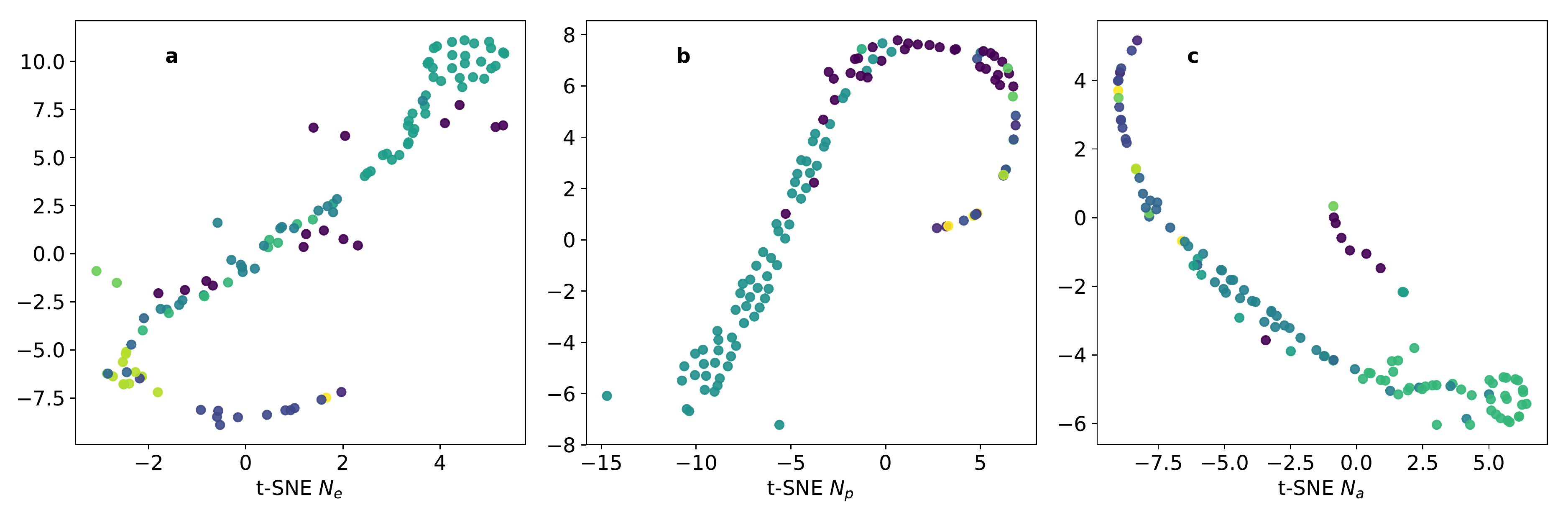}
\caption{
The t-distributed stochastic neighbour embedding (t-SNE)~\cite{Maaten2008} of 3-vector of Wikimedia projects onto 2-dimensional space (see Supplementary Methods for the details). \textbf{a}, embedded vector from 3-vector of number of edits $N_e$. \textbf{b}, embedded vector from 3-vector of number of editors $N_p$. \textbf{c}, embedded vector from 3-vector of number of articles $N_a$. Each point corresponds to a certain language in Wikimedia project, and its colour corresponds to a cluster to which the language belongs. Each language is clustered by Dirichlet Process Gaussian Mixture Model~\cite{Blie2006} with a parameter $\gamma_0 = 10^{-2}$, yet the result is stable within the range of $\gamma_0 \in [10^{-2}, 10^2]$.
}
\label{fig:wiki_cluster_3vec}
\end{figure*}

\pagebreak
\null
\begin{figure*}[!ht]
\includegraphics[width=\textwidth]{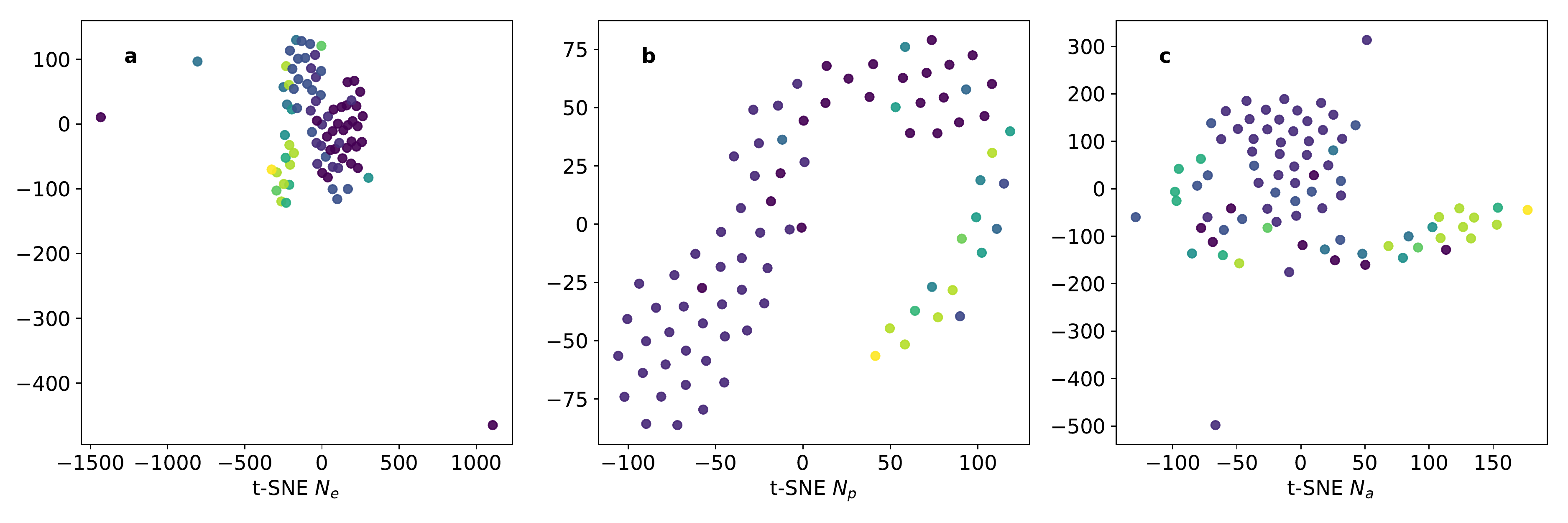}
\caption{The t-distributed stochastic neighbour embedding (t-SNE)~\cite{Maaten2008} of 4-vector of Wikimedia projects onto 2-dimensional space (see Supplementary Methods for the details). \textbf{a}, embedded vector from 4-vector of number of edits $N_e$. \textbf{b}, embedded vector from 4-vector of number of editors $N_p$. \textbf{c}, embedded vector from 4-vector of number of articles $N_a$. Each point corresponds to a certain language in Wikimedia project, and its colour corresponds to a cluster to which the language belongs. Each language is clustered by Dirichlet Process Gaussian Mixture Model~\cite{Blie2006} with a parameter $\gamma_0 = 10^{-2}$, yet the result is stable within the range of $\gamma_0 \in [10^{-2}, 10^2]$.}\label{fig:wiki_cluster_4vec}
\end{figure*}
\pagebreak
\null
\begin{figure*}[!ht]
\includegraphics[width=\textwidth]{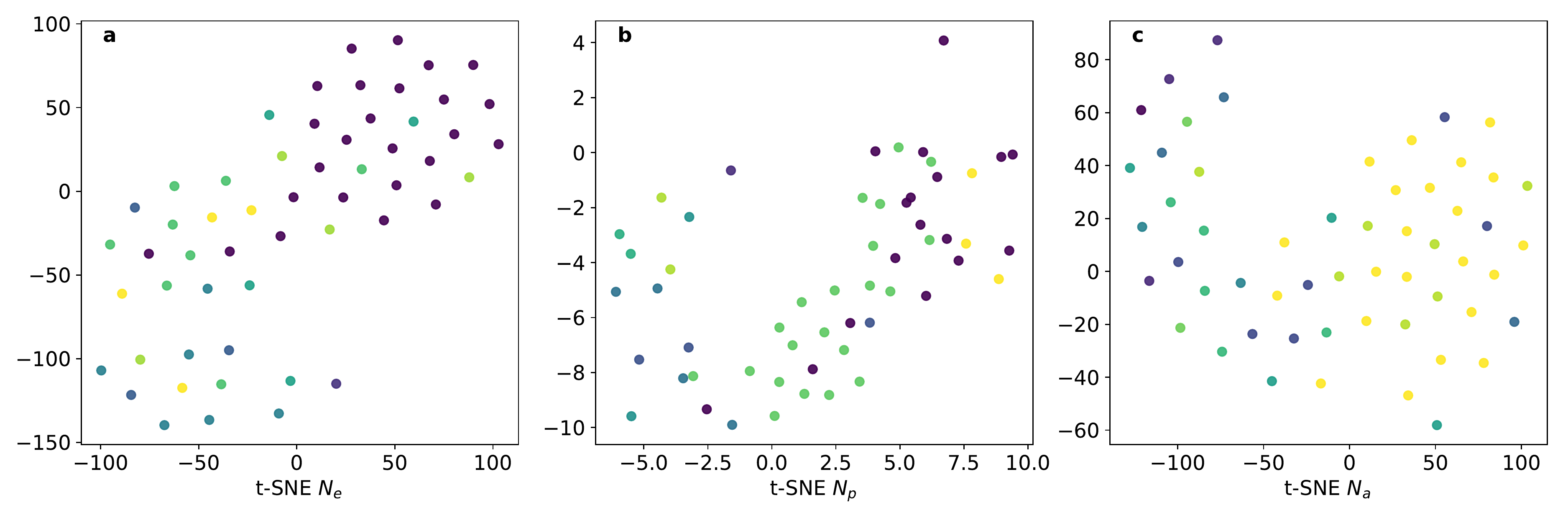}
\caption{The t-distributed stochastic neighbour embedding (t-SNE)~\cite{Maaten2008} of 5-vector of Wikimedia projects onto 2-dimensional space (see Supplementary Methods for the details). \textbf{a}, embedded vector from 5-vector of number of edits $N_e$. \textbf{b}, embedded vector from 5-vector of number of editors $N_p$. \textbf{c}, embedded vector from 5-vector of number of articles $N_a$. Each point corresponds to a certain language in Wikimedia project, and its colour corresponds to a cluster to which the language belongs. Each language is clustered by Dirichlet Process Gaussian Mixture Model~\cite{Blie2006} with a parameter $\gamma_0 = 10^{-2}$, yet the result is stable within the range of $\gamma_0 \in [10^{-2}, 10^2]$. 
}
\label{fig:wiki_cluster_5vec}
\end{figure*}

\pagebreak
\null
\begin{figure*}[!ht]
\includegraphics[width=\textwidth]{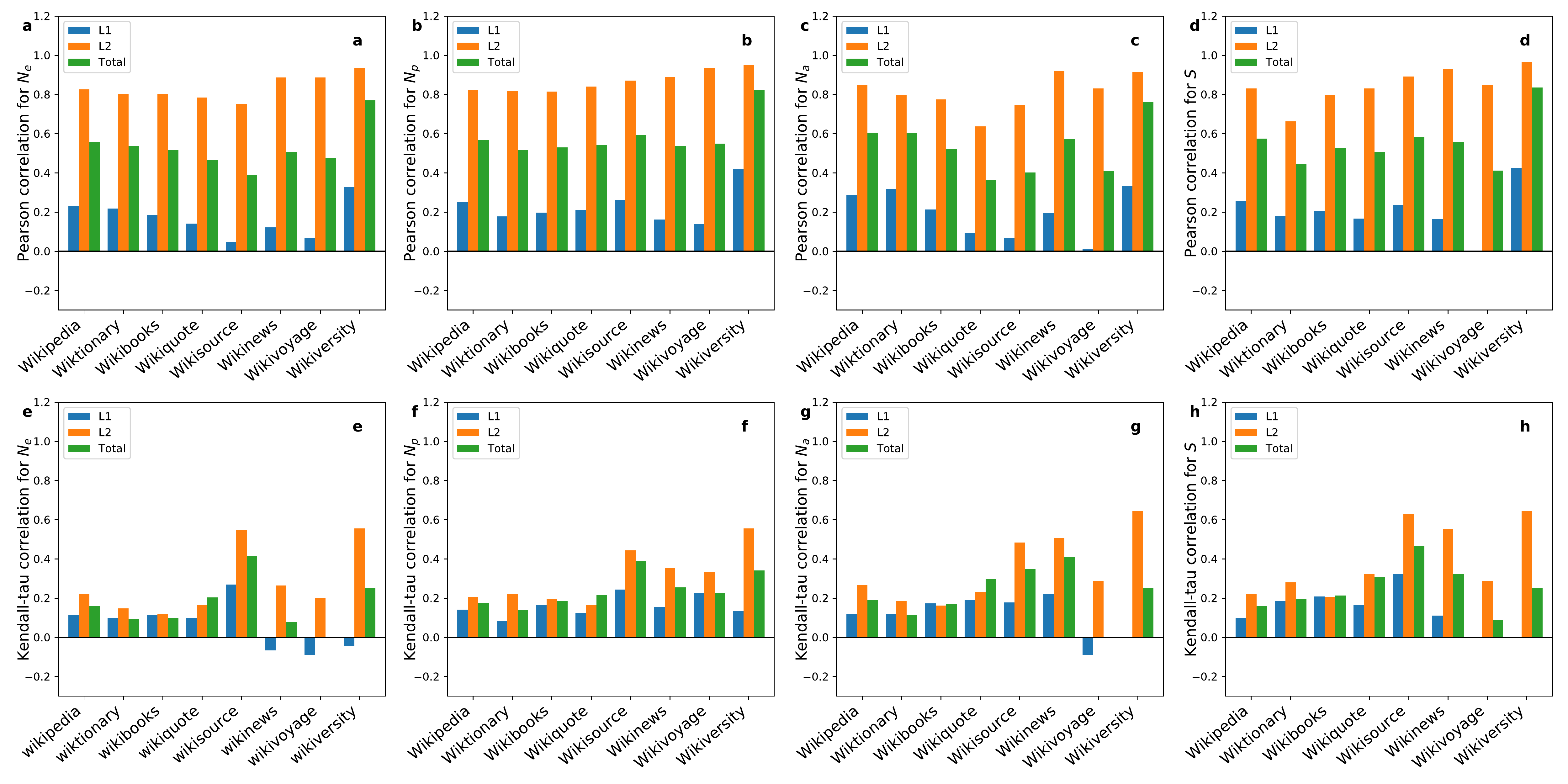}
\caption{The correlations of total number of speakers as functions of measures of Wikimedia project: (\textbf{a,e}), the number of edits $N_e$, (\textbf{b,f}), the number of editors $N_p$, \textbf{(c,g)}, the number of articles $N_a$, and (\textbf{d,h}), the total volume of texts $S$ (in the unit of bytes). For (\textbf{a--h}) colours of the bar indicate number of the language's level 1 speakers (blue bars; as a native language), level 2 speakers (orange bars; as a secondary language), and total number of speakers (green bars). All of the plots are drawn under the conditions as follows: 1) the number of speakers are estimated by 2017 edition of Ethnologue~\cite{Ethnologue2017}, 2) at least 50 million speakers should be in total, and 3) languages with no information on level 2 speakers are assumed to not having level 2 speakers. As a result, we used 143 Wikimedia projects in 26 distinct languages in total. Please note that the estimation of language users is generally not reliable, because they were collected from different sources and dates \cite{Paolillo2006}. Both the Pearson correlation coefficient (\textbf{a--d}) and the Kendall-tau rank correlation coefficient (\textbf{e--h}) are used to quantify interplays between measures, to exclude possible biases when we only observe a single measure.}
\label{fig:languser}
\end{figure*}

\pagebreak
\null
\begin{figure*}[!ht]
\includegraphics[width=\textwidth]{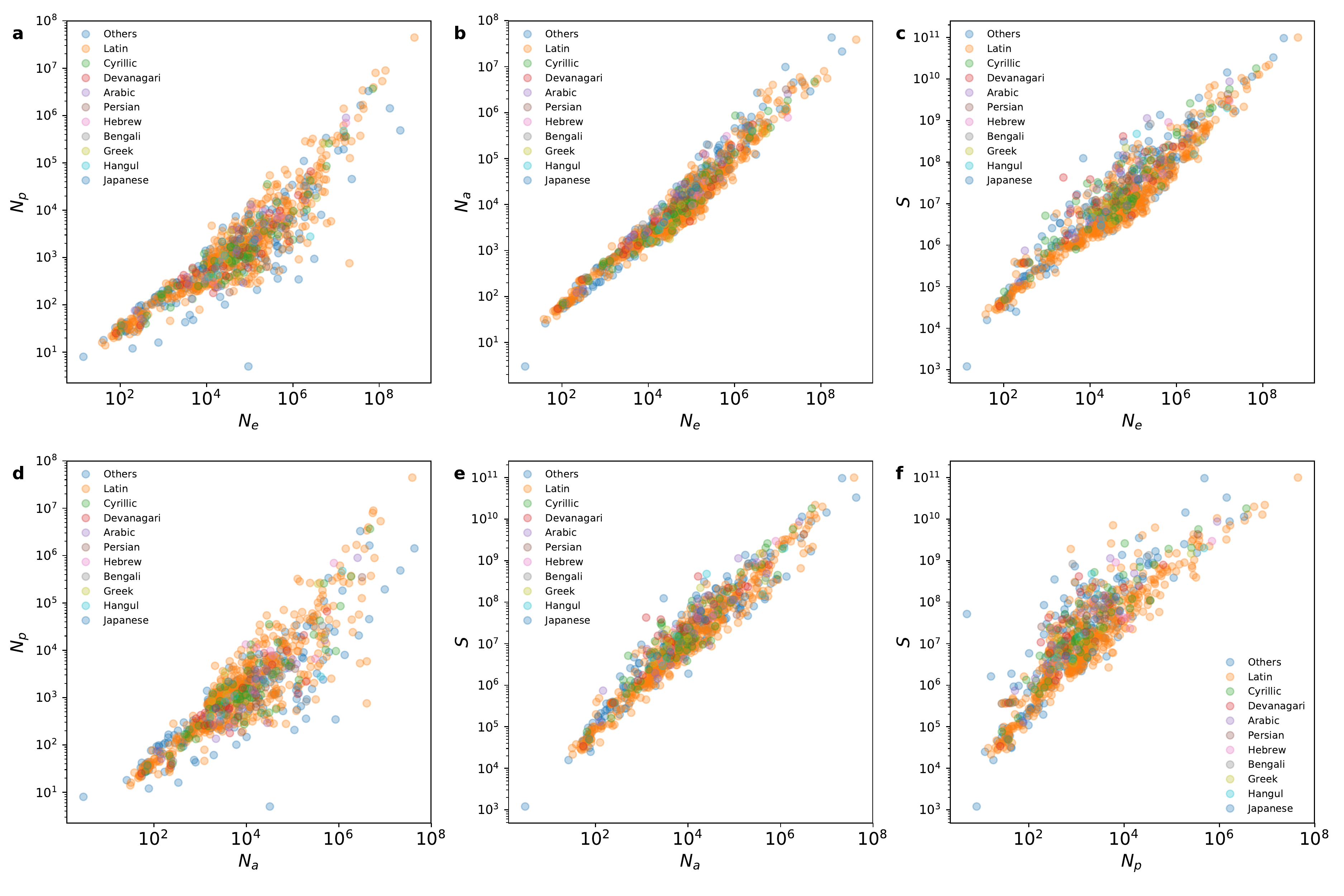}
\caption{The correlations between number of edits $N_e$, number of editors $N_p$, number of articles $N_a$, and total size of the data set $S$. The colours represent the script with which the corresponding Wikimedia projects are written.}
\label{fig:wiki_script}
\end{figure*}

\pagebreak
\begin{figure*}[!ht]
\includegraphics[width=\textwidth]{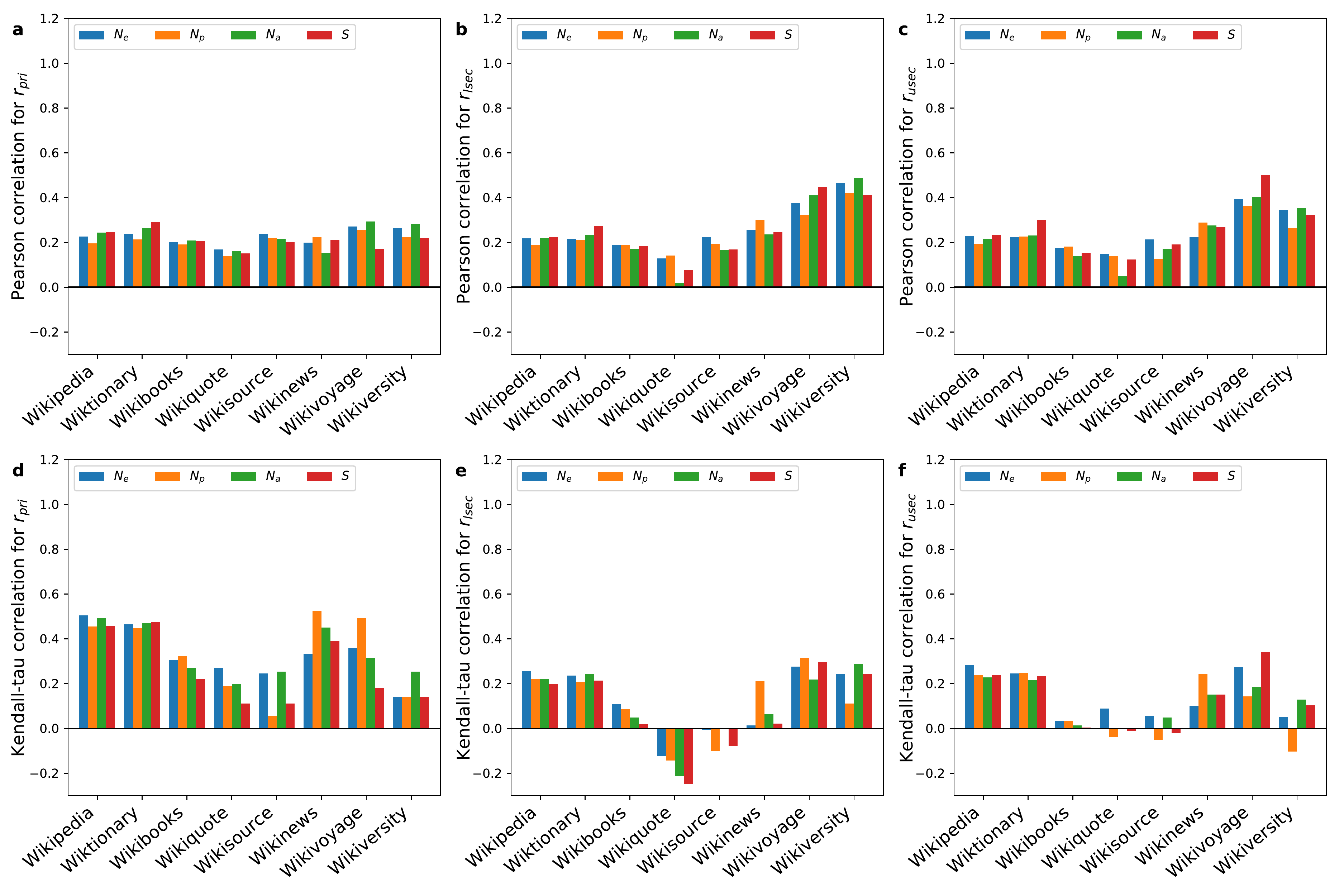}
\caption{The correlations between measures of different Wikimedia projects and education levels of the corresponding country (See Supplementary Methods for details). Education levels are collected from UNESCO Institute for Statistics~\cite{UIS} \textbf{a,d}, The correlation coefficients on measures of Wikimedia projects for attainment rate of a primary school of the corresponding country. \textbf{b,e}, Correlation coefficients on measures of Wikimedia projects for attainment rate of a lower secondary school of the corresponding country. \textbf{c,f}, The correlation coefficients on measures of Wikimedia projects for attainment rate of a upper secondary school of the corresponding country.  For panels (\textbf{a--f}), the colours of the bars indicate the number of edits $N_e$ (blue bars), the number of editors $N_p$ (orange bars), the number of articles $N_a$ (green bars), and the total volume of texts $S$ (red bars). Both the Pearson correlation coefficient (\textbf{a--c}) and the Kendall-tau rank correlation coefficient (\textbf{d--f}) is used to quantify interplays between measures, to exclude possible biases when we only observe a single measure.}
\label{fig:education}
\end{figure*}

\pagebreak
\null
\begin{figure*}[!ht]
\includegraphics[width=\textwidth]{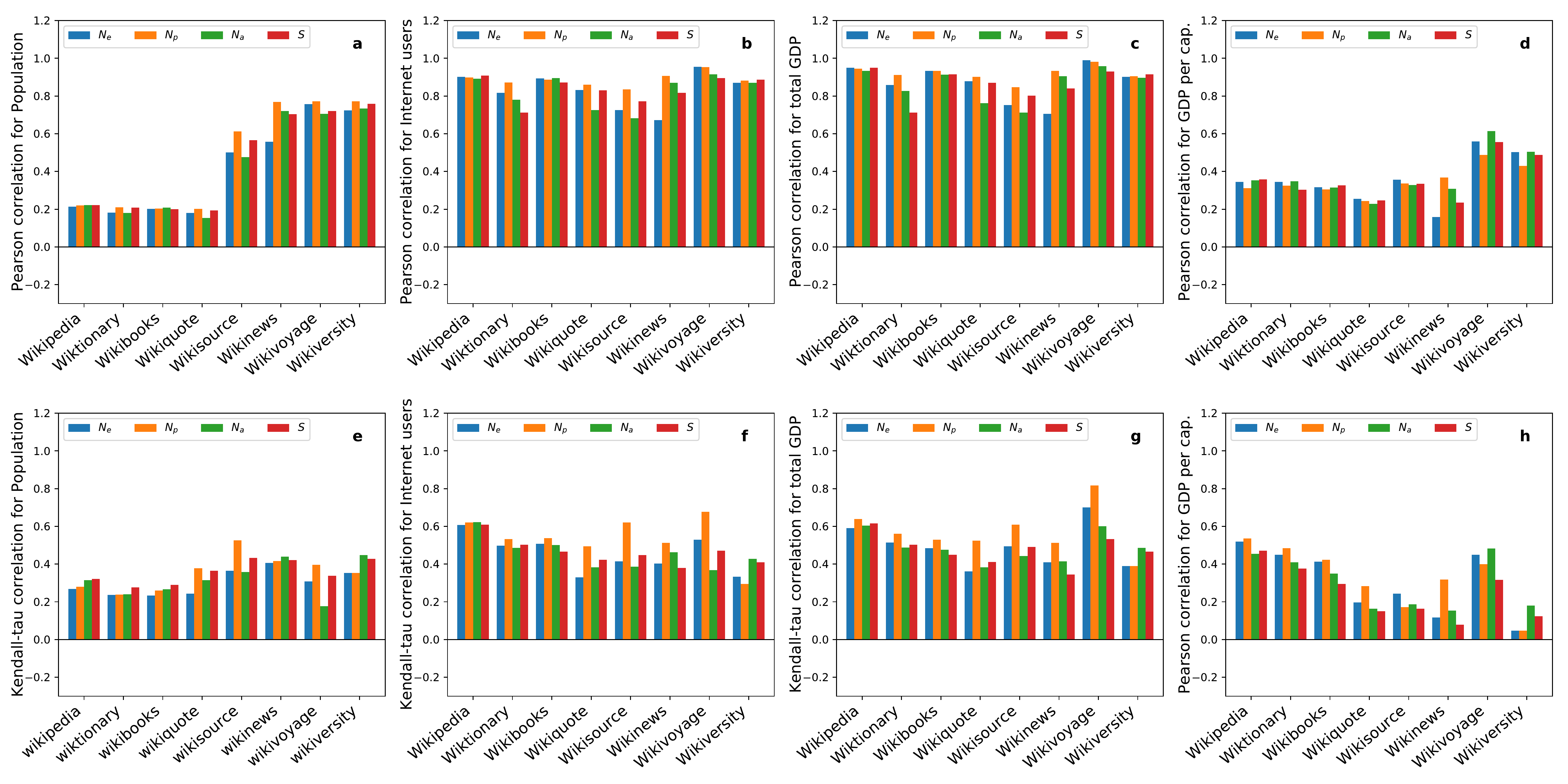}
\caption{The correlations between measures of different Wikimedia projects and scocio-economic indicators of the corresponding country (See Supplementary Methods for details). Population and numbers of Internet users are retrieved from Central Intelligence Agency (CIA) world FACT Book~\cite{CIA16}, and GDP measures are collected from UNESCO Institute for Statistics~\cite{UIS}. \textbf{a,e}, The correlation coefficients on measures of Wikimedia projects for the population of the corresponding country. \textbf{b,f}, The correlation coefficients on measures of Wikimedia projects for the number of Internet users of the corresponding country. \textbf{c,g}, Correlation coefficients on measures of Wikimedia projects for the GDP at market prices (in constant 2010 US\$) of the corresponding country. \textbf{d,h}, The correlation coefficients on measures of Wikimedia projects for the GDP per capita, or PPP (in constant 2011 international \$) of the corresponding country. For panels (\textbf{a--h}), the colours of the bars indicate the number of edits $N_e$ (blue bars), the number of editors $N_p$ (orange bars), the number of articles $N_a$ (green bars), and the total volume of texts $S$ (red bars). Both the Pearson correlation coefficient (\textbf{a--d}) and the Kendall-tau rank correlation coefficient (\textbf{e--h}) is used to quantify interplays between measures, to exclude possible biases when we only observe a single measure.}
\label{fig:economy}
\end{figure*}

\pagebreak
\begin{figure*}[!ht]
\includegraphics[width=\textwidth]{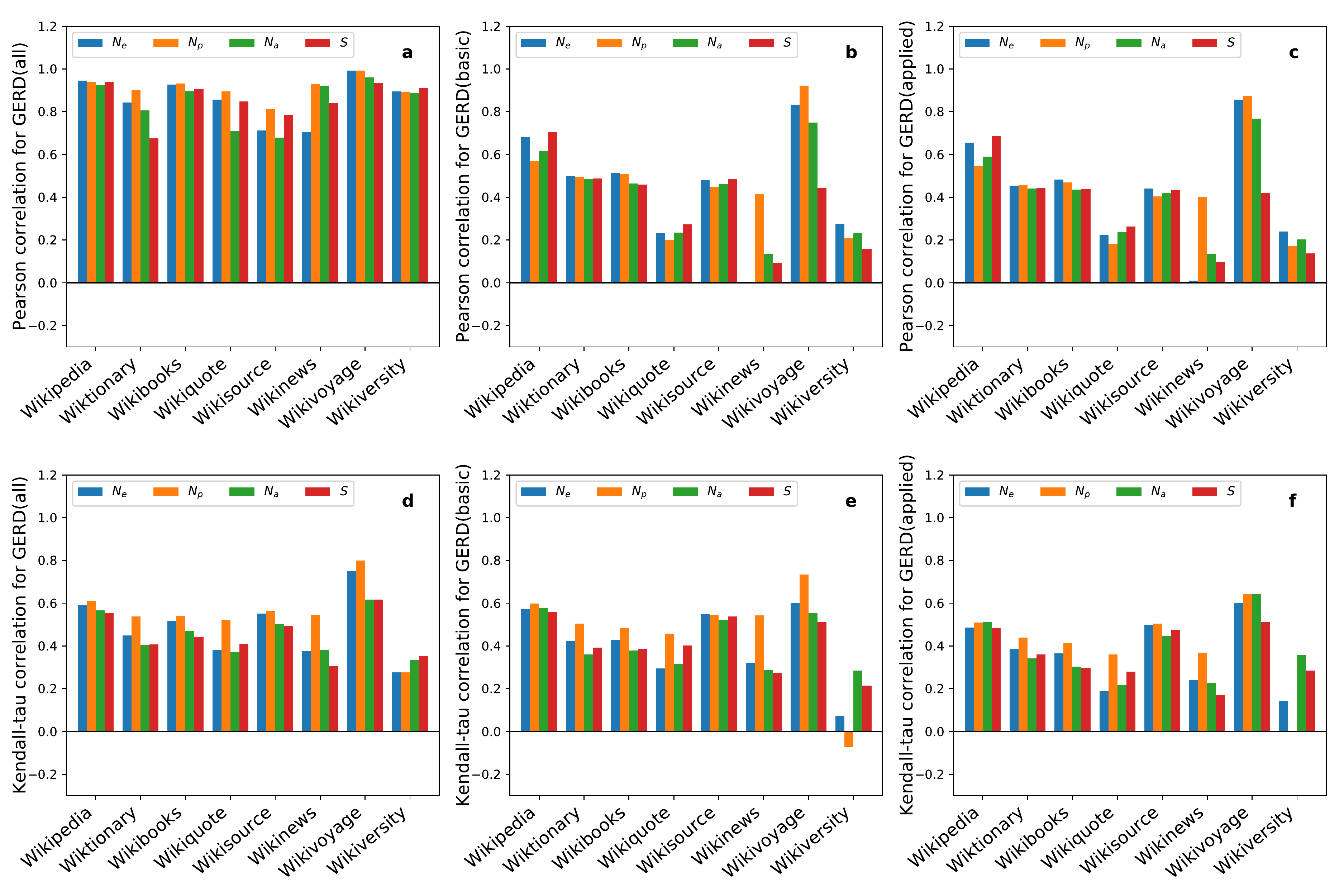}
\caption{The correlations between measures of different Wikimedia projects and gross domestic expenditure on R\&D (GERD) of the corresponding country collected from UNESCO Institute for Statistics~\cite{UIS} (See Supplementary Methods for details). \textbf{a,d}, The correlation coefficients on measures of Wikimedia projects for the total gross domestic expenditure on R\&D of the corresponding country. \textbf{b,e}, The correlation coefficients on measures of Wikimedia projects for the gross domestic expenditure on R\&D for basic research of the corresponding country. \textbf{c,f}, The correlation coefficients on measures of Wikimedia projects for the gross domestic expenditure on R\&D for applied research of the corresponding country. For (\textbf{a--f}) colours of the bar indicate the number of edits $N_e$ (blue bars), the number of editors $N_p$ (orange bars), the number of articles $N_a$ (green bars), and the total volume of texts $S$ (red bars). Both the Pearson correlation coefficient (\textbf{a--c}) and the Kendall-tau rank correlation coefficient (\textbf{d--f}) is used to quantify interplays between measures, to exclude possible biases when we only observe a single measure.}
\label{fig:rndinvest}
\end{figure*}

\pagebreak
\null
\begin{figure*}[!ht]
\includegraphics[width=\textwidth]{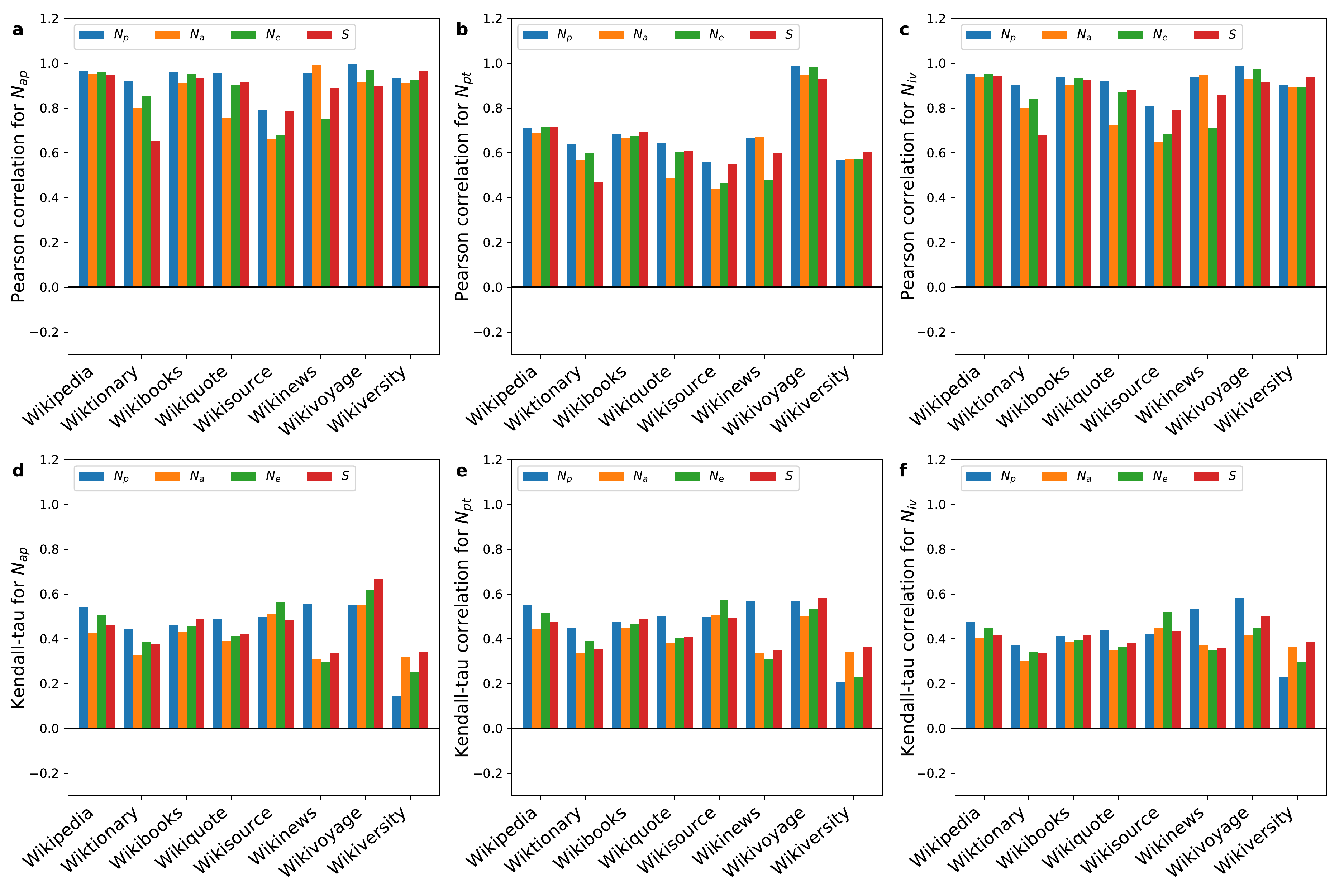}
\caption{The correlations between measures of different Wikimedia projects and measures of patent applications (see Supplementary Methods for details). \textbf{a,d}, The correlation coefficients on measures of Wikimedia projects for the number of patent applicants $N_{ap}$ \textbf{b,e}, The correlation coefficients on measures of Wikimedia projects for number of patents $N_{pt}$. \textbf{c,f}, The correlation coefficients on measures of Wikimedia projects for the number of patent inventors $N_{iv}$. For panels (\textbf{a--f}), the colours of the bars indicate the number of edits $N_e$ (blue bars), the number of editors $N_p$ (orange bars), the number of articles $N_a$ (green bars), and the total volume of texts $S$ (red bars). Both the Pearson correlation coefficient (\textbf{a--c}) and the Kendall-tau rank correlation coefficient (\textbf{d--f}) is used to quantify interplays between measures, to exclude possible biases when we only observe a single measure.
}
\label{fig:patent}
\end{figure*}

\pagebreak
\null
\begin{figure*}[!ht]
\includegraphics[width=0.67\textwidth]{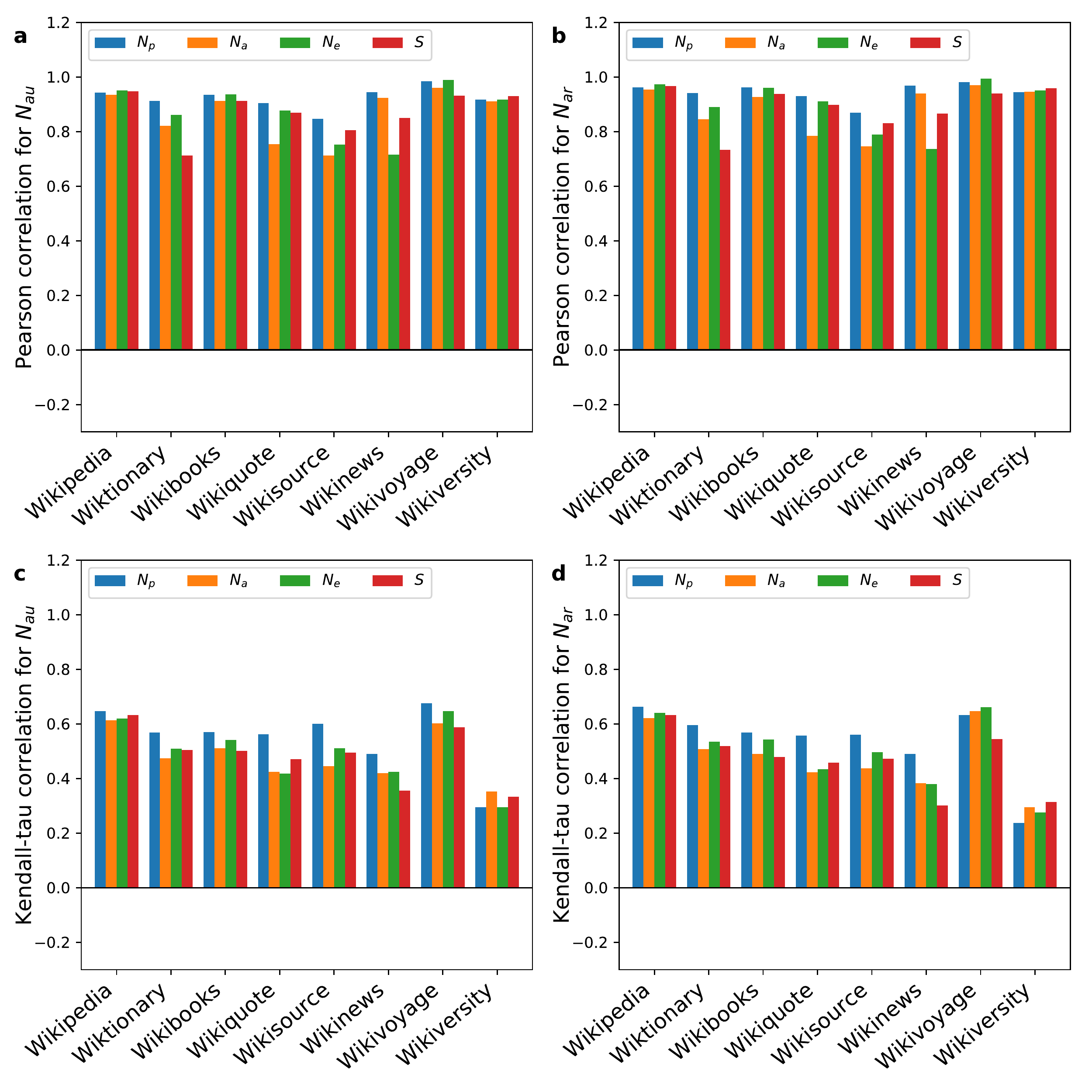}
\caption{The correlations between measures of different Wikimedia projects and measures of research articles (see Supplementary Methods for details). \textbf{a,c}, The correlation coefficients on measures of Wikimedia projects for the number of authors $N_{au}$. \textbf{b,d}, The correlation coefficients on measures of Wikimedia projects for number of research articles $N_{ar}$. For panels (\textbf{a--d}), the colours of the bar indicate the number of edits $N_e$ (blue bars), the number of editors $N_p$ (orange bars), the number of articles $N_a$ (green bars), and the total volume of texts $S$ (red bars). Both the Pearson correlation coefficient (\textbf{a,b}) and the Kendall-tau rank correlation coefficient (\textbf{c,d}) is used to quantify interplays between measures, to exclude possible biases when we only observe a single measure.
}
\label{fig:paper}
\end{figure*}

\pagebreak
\null
\begin{figure*}[!ht]
\includegraphics{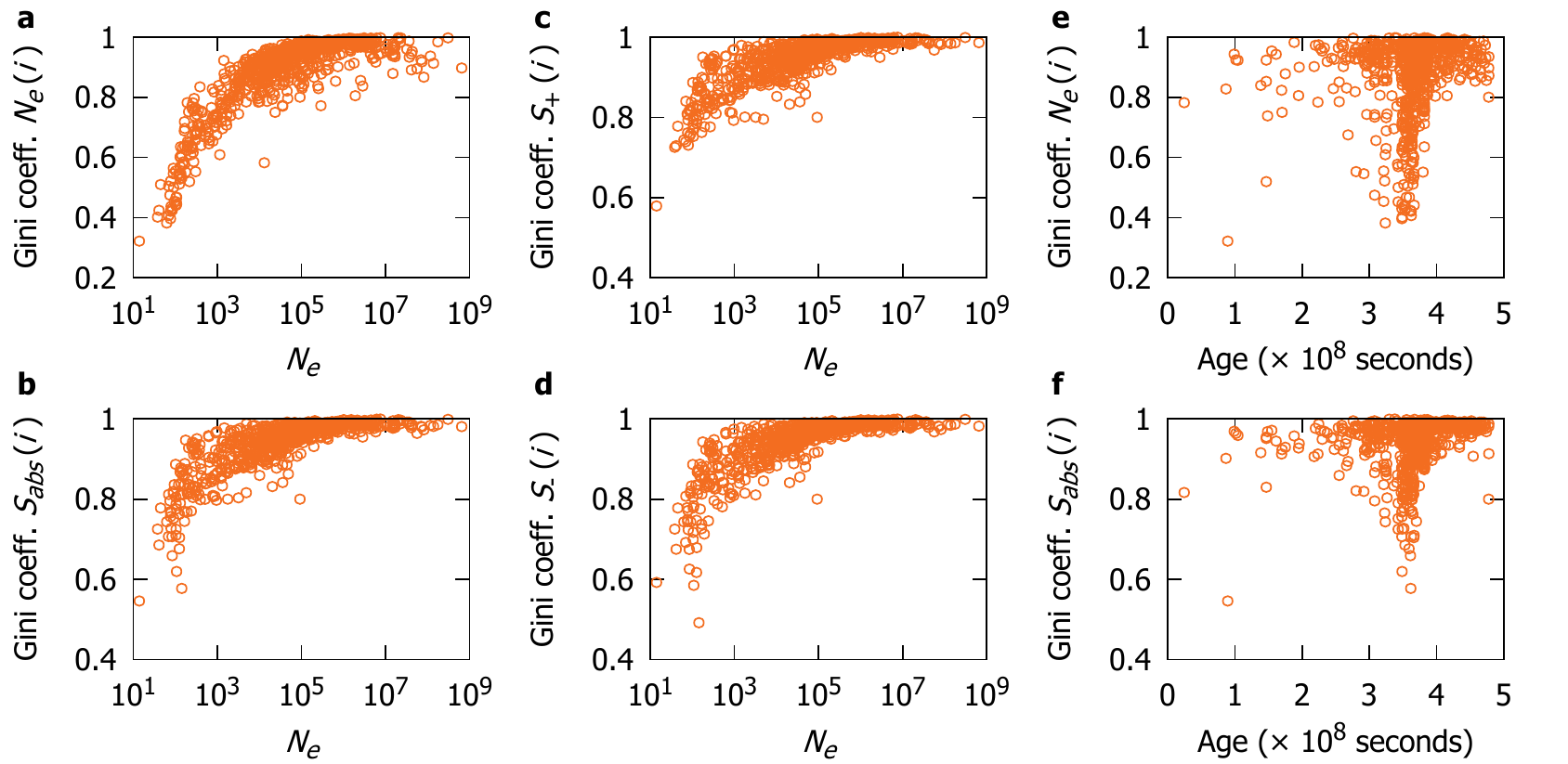}
\caption{The Gini coefficient of Wikimedia projects as functions of the number of edits (\textbf{a--d}) and the real time (\textbf{e, f}). \textbf{a}, The Gini coefficient for numbers of edits performed by each editor. \textbf{b}, The average Gini coefficient for total sums of absolute amount of editing (in the unit of bytes) performed by each editor. \textbf{c}, The average Gini coefficient for the absolute number of sums of incremental change in edits performed by each editor. \textbf{d}, The average Gini coefficient for the absolute number of sums of decremental change in edits performed by each editor. \textbf{e}, The Gini coefficient for numbers of edits performed by each editor. \textbf{f}, The average Gini coefficient for total sums of absolute amount of editing (in the unit of bytes) performed by each editor.}
\label{fig:wiki_gini}
\end{figure*}

\pagebreak
\null
\begin{figure*}[!ht]
\includegraphics[width=0.67\textwidth]{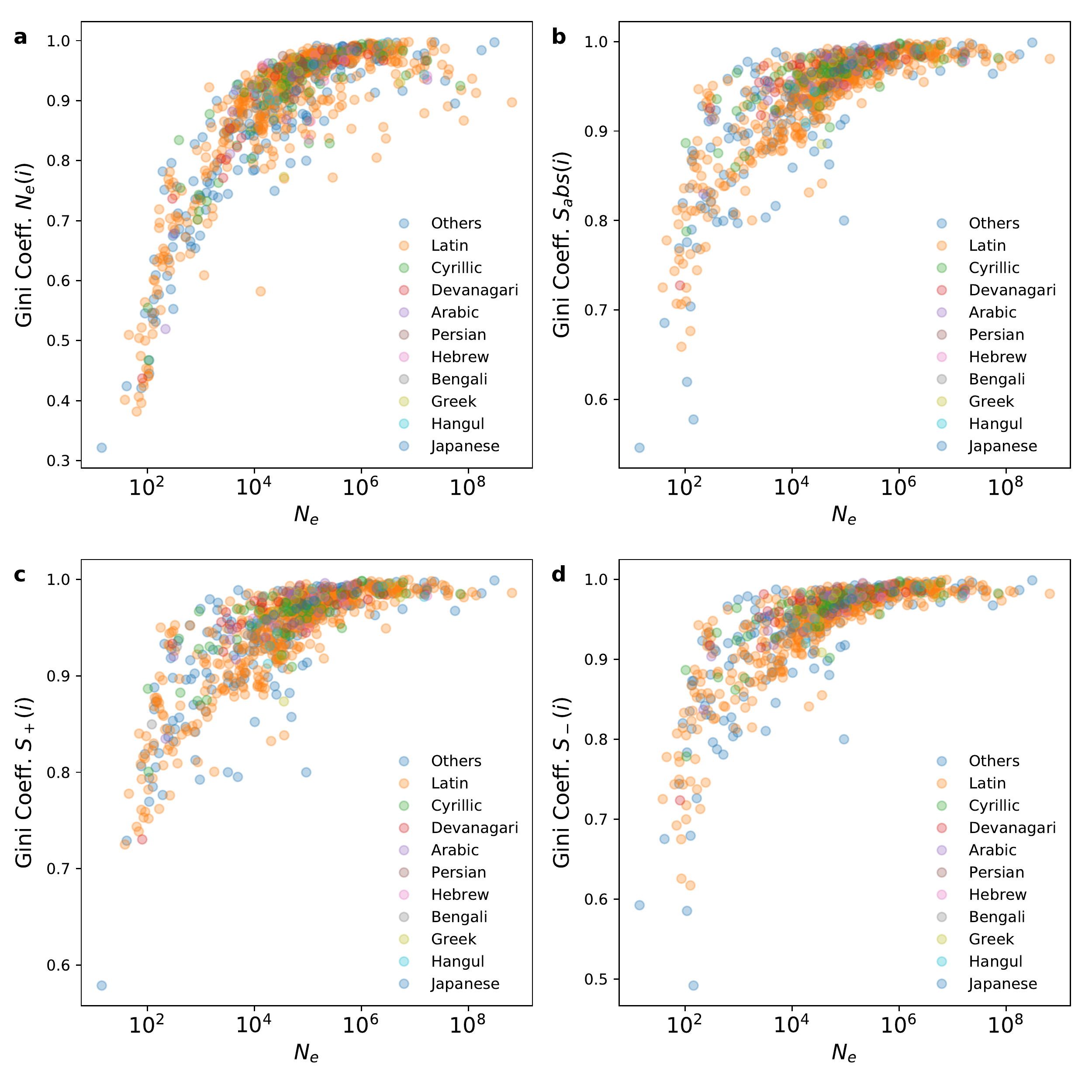}
\caption{The Gini coefficient of Wikimedia projects as functions of the number of edits. \textbf{a}, The Gini coefficient for numbers of edits performed by each editor. \textbf{b}, The average Gini coefficient for total sums of absolute amount of editing (in the unit of bytes) performed by each editor. \textbf{c}, The average Gini coefficient for the absolute number of sums of incremental change in edits performed by each editor. \textbf{d}, The average Gini coefficient for the absolute number of sums of decremental change in edits performed by each editor. The colours represent the script with which the corresponding Wikimedia projects are written.}
\label{fig:wiki_script_gini}
\end{figure*}

\pagebreak
\null
\begin{figure*}[!ht]
\includegraphics[width=0.67\textwidth]{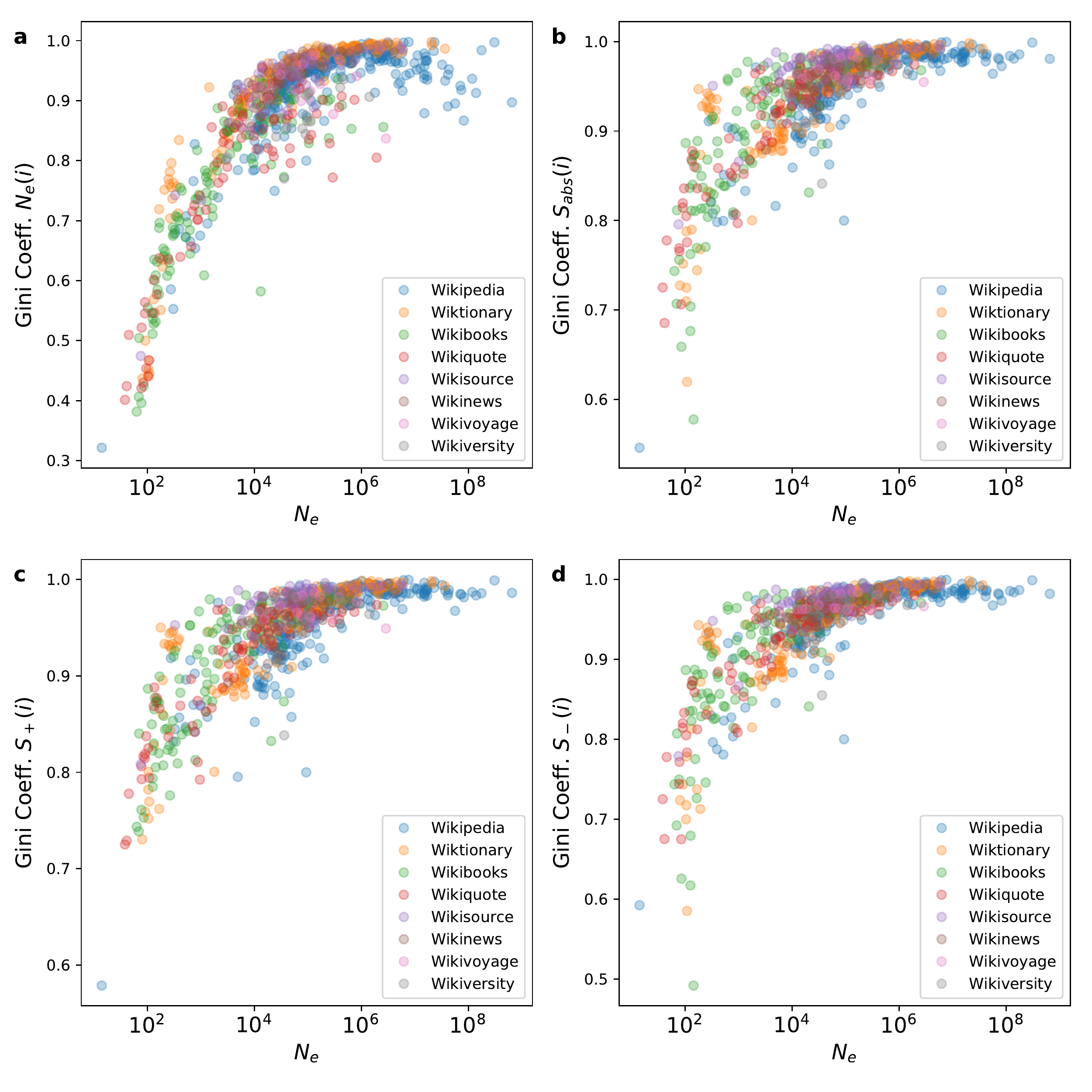}
\caption{The Gini coefficient of Wikimedia projects as functions of the number of edits. \textbf{a}, The Gini coefficient for numbers of edits performed by each editor. \textbf{b}, The average Gini coefficient for total sums of absolute amount of editing (in the unit of bytes) performed by each editor. \textbf{c}, The average Gini coefficient for the absolute number of sums of incremental change in edits performed by each editor. \textbf{d}, The average Gini coefficient for the absolute number of sums of decremental change in edits performed by each editor. The colours represent the types of Wikimedia projects.}
\label{fig:wiki_type_gini}
\end{figure*}

\pagebreak
\null
\begin{figure*}[!ht]
\includegraphics[width=\textwidth]{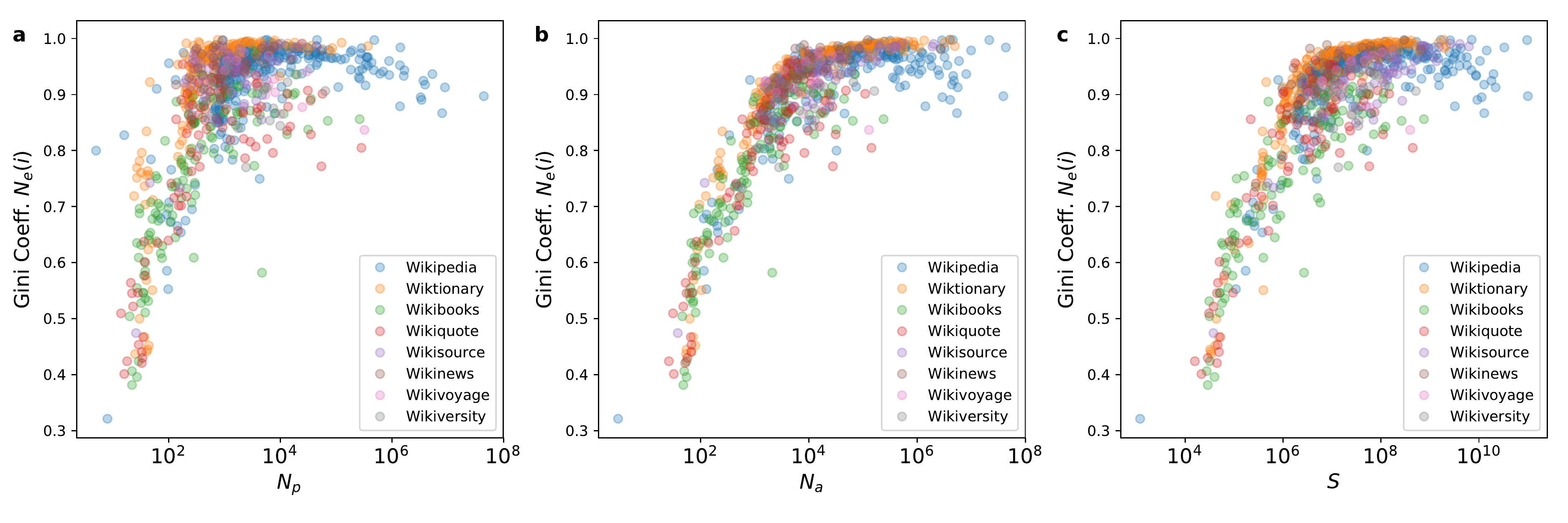}
\caption{The Gini coefficient of Wikimedia projects as functions of the characteristic measures of the projects. \textbf{a}, The Gini coefficient for numbers of edits performed by each editor as a function of the number of editors $N_p$. \textbf{b}, The Gini coefficient for numbers of edits performed by each editor as a function of the number of articles $N_a$.  \textbf{c}, The Gini coefficient for numbers of edits performed by each editor as a function of the total size of the data set $S$. The colours represent the types of Wikimedia projects.}
\label{fig:wiki_type_gini2}
\end{figure*}

\pagebreak
\null
\begin{figure*}[!ht]
\includegraphics{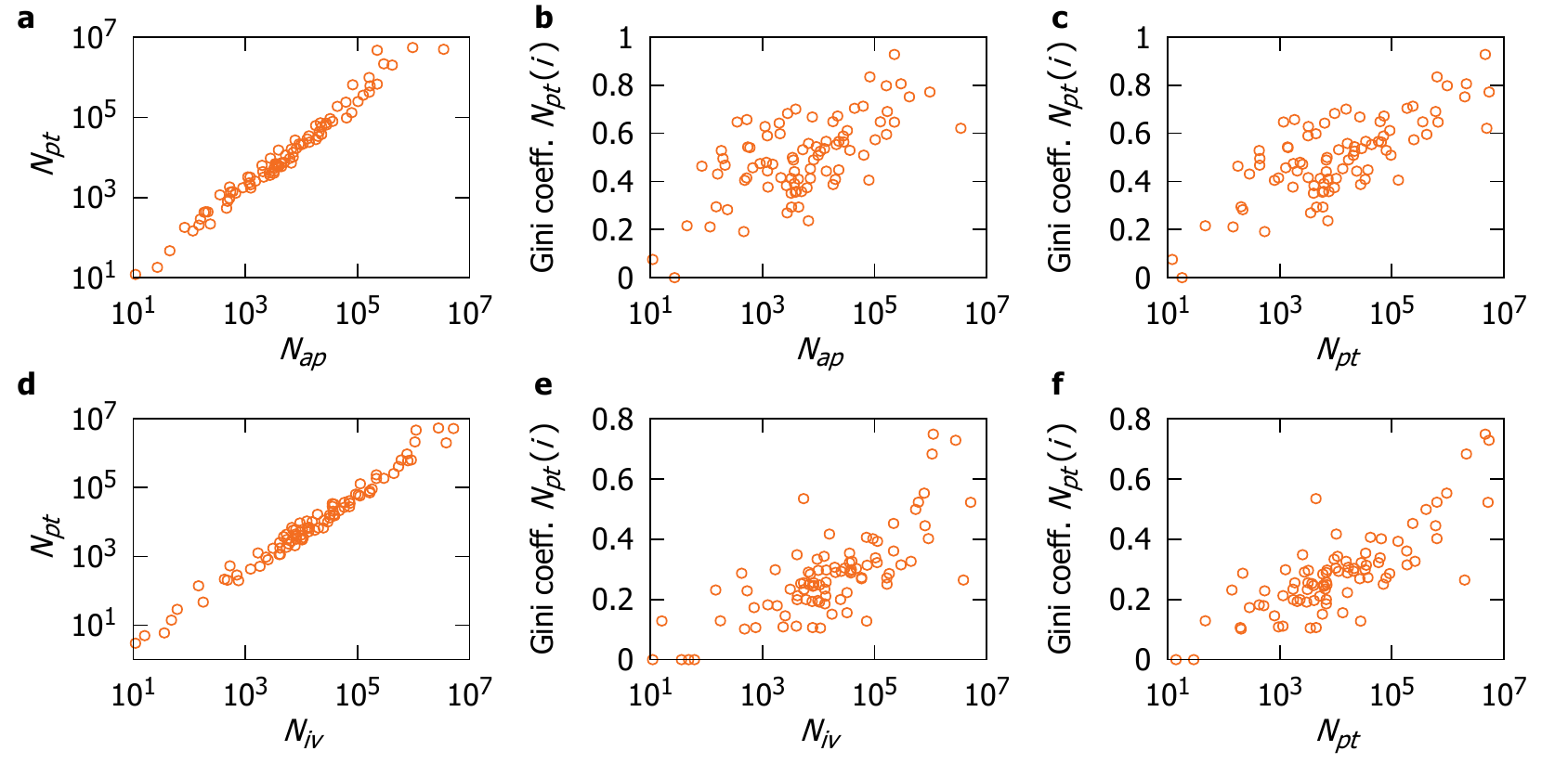}
\caption{The correlations and the Gini coefficients of patent applications as functions of number of patents $N_{pt}$ for number of patent applicants $N_{ap}$ (\textbf{a--c}) and patent inventors $N_{iv}$ (\textbf{d--f}). \textbf{a}, The numbers of patent as functions of the number of patent applicants. \textbf{b}, The Gini coefficient for the numbers of patents filed by each applicant, as functions of the number of patent applicants. \textbf{c}, The Gini coefficient for numbers of patents filed by each applicant, as functions of the number of patents. \textbf{d}, The numbers of patent as functions of the number of patent inventors. \textbf{e}, The Gini coefficient for the numbers of patents filed by each inventor, as functions of the number of patent inventors. \textbf{c}, The Gini coefficient for numbers of patents filed by each inventor, as functions of the number of patents. For panels (\textbf{a--f}), each data point corresponds to a patent office that the patent initially filed. For the analysis, we use spring 2017 edition of European Patent Office (EPO) Worldwide Patent Statistical Database (PATSTAT), which contains metadata of patents from 91 national and international patent offices. Inventors and applicants are identified as a person IDs by EPO and we neglected the persons who are not identified by EPO (see Supplementary Methods for details).}
\label{fig:patent_gini}
\end{figure*}

\pagebreak
\null
\begin{figure*}[!ht]
\includegraphics{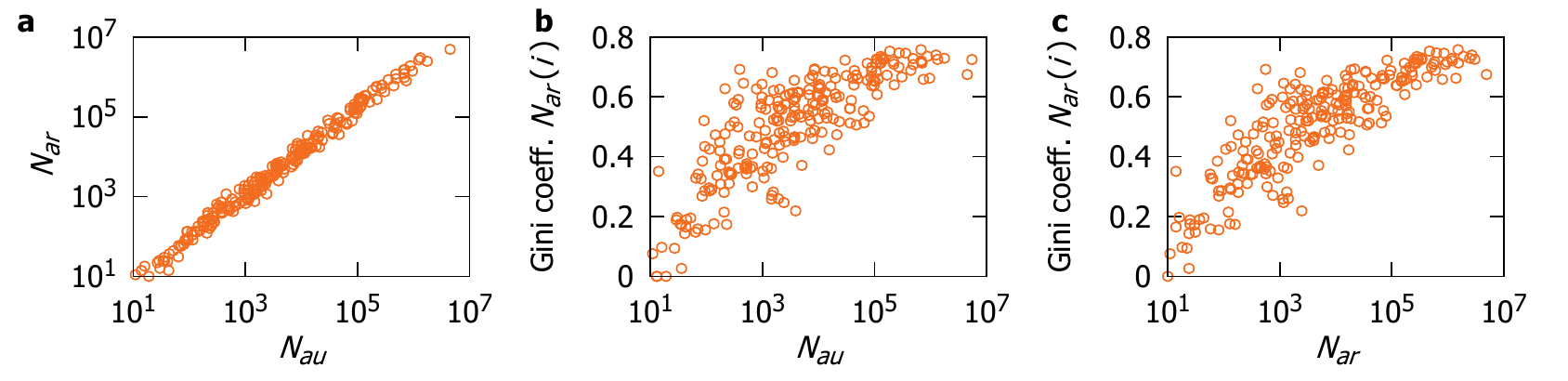}
\caption{The correlations and the Gini coefficients of number of authors as functions of number of research articles $N_{ar}$ for number of authors $N_{au}$. \textbf{a}, Numbers of research articles as functions of the number of authors. \textbf{b}, The Gini coefficient for numbers of research articles published by each authors, as functions of the number of authors. \textbf{c}, The Gini coefficient for numbers of research articles published by each author, as functions of the number of research articles. Each data point corresponds to a nationality to which the affiliation of the author belongs. For the analysis, we use August 2017 edition of SCOPUS XML CUSTOM DATA, which contains meta data of worldwide research outputs. Authors and their affiliations are identified as an author IDs and affiliation IDs given by SCOPUS and we neglected the authors who are not identified by SCOPUS (see Supplementary Methods for details).}
\label{fig:paper_gini}
\end{figure*}

\pagebreak
\null
\begin{figure*}[!ht]
\includegraphics[width=0.7\textwidth]{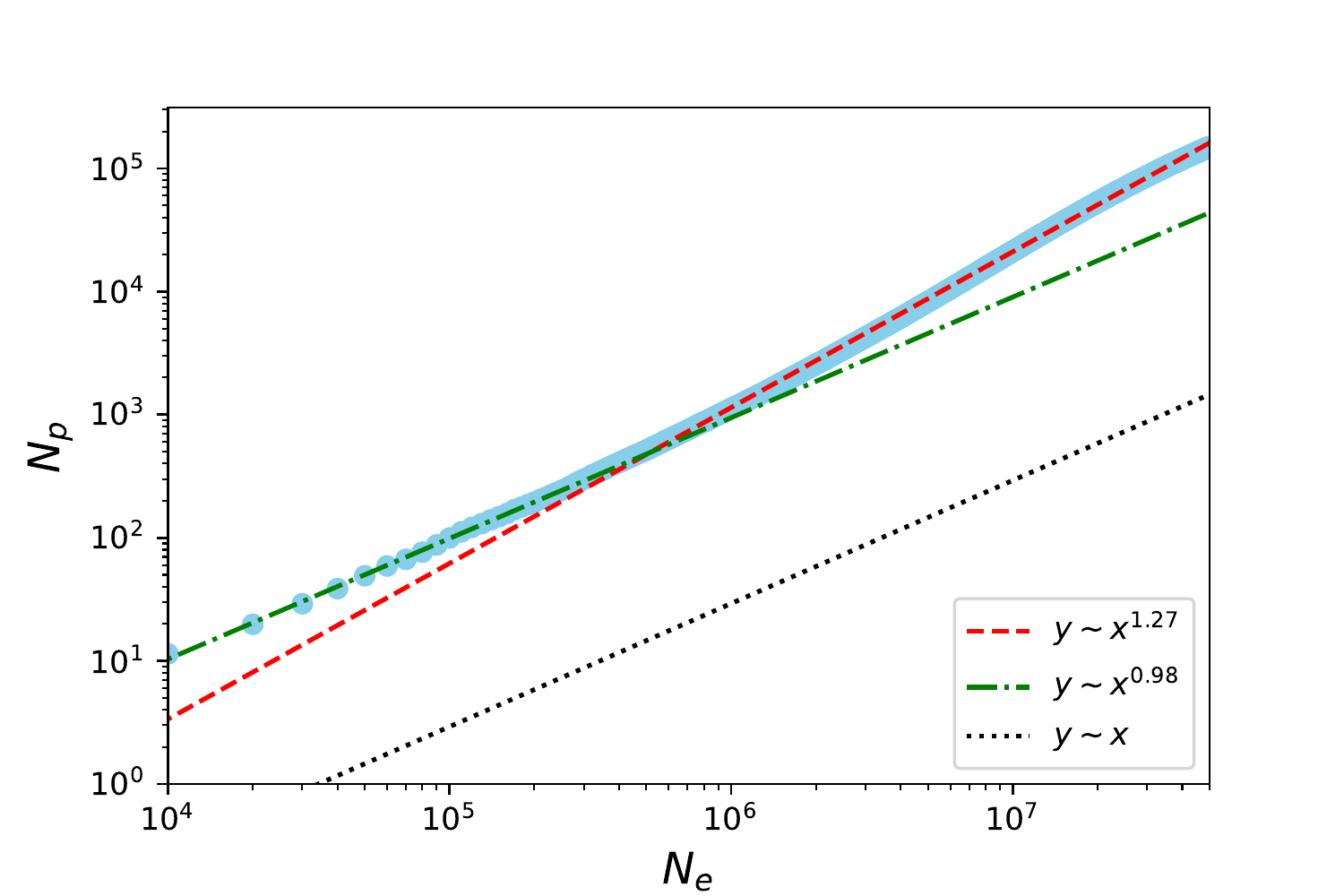}
\caption{The scaling relation between number of edits $N_e$ and number of editors $N_p$ of the model result. The relation is characterized by two simple power-law growth form of $y \sim x^\lambda$. For the initial stage $N < 10^6$, the number of editors increases sublinearly with the exponent $\lambda \simeq 0.98$, whereas the number of editors increases superlinearly with the exponent $\lambda \simeq 1.27$ for later ($N > 10^6$). We use the following parameters with $10$ realizations: $b = 0.0001$, $k = 0.4$, $\tau \to \infty$, and $r = 0.01$.}  
\label{fig:model_npne_0_4}
\end{figure*}

\pagebreak
\null
\begin{figure*}[!ht]
\includegraphics[width=0.7\textwidth]{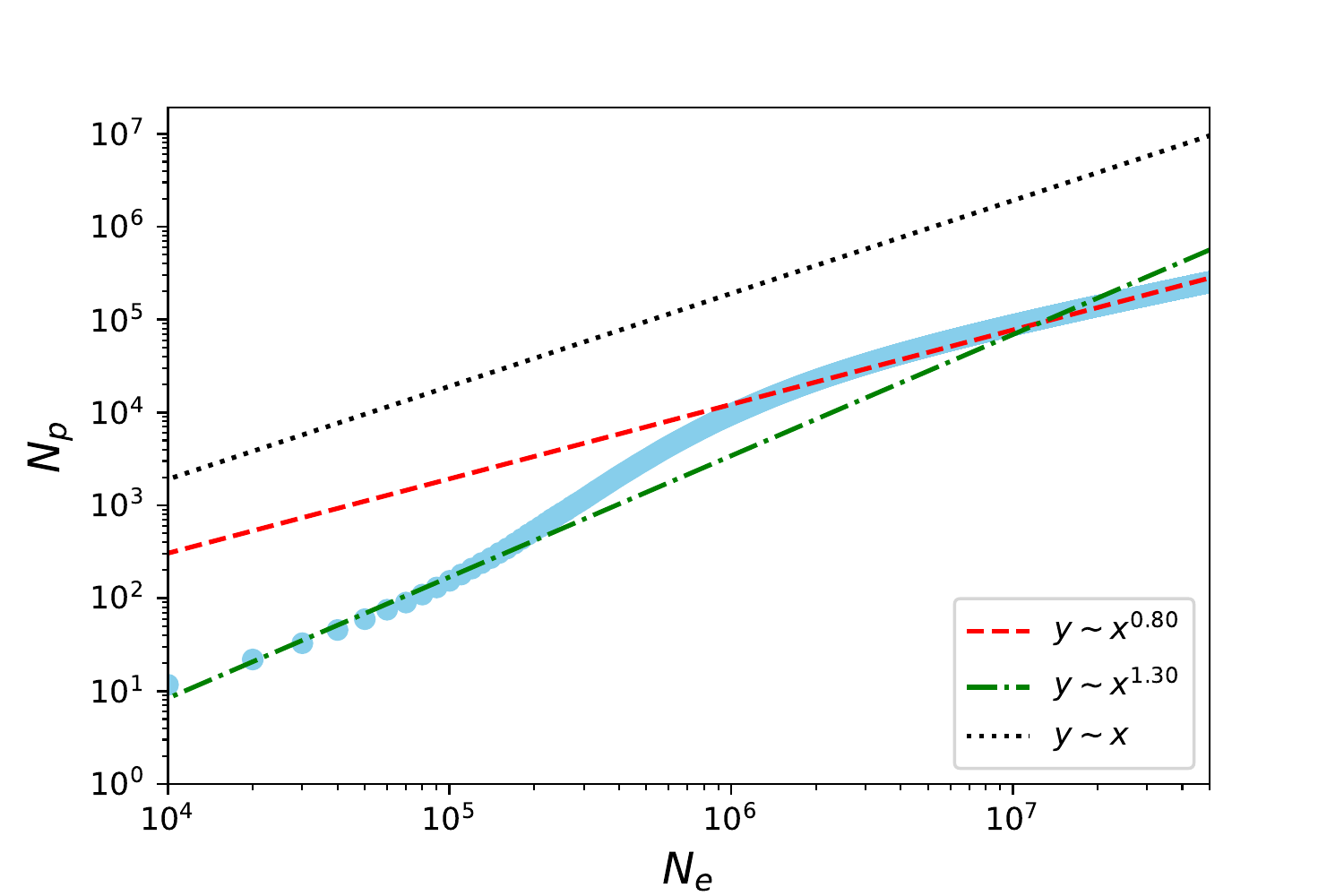}
\caption{The scaling relation between number of edits $N_e$ and number of editors $N_p$ of the model result. The relation is characterized by two simple power-law growth form of $y \sim x^\lambda$. For the initial stage $N < 10^6$, the number of editors increases superlinearly with an exponent $\lambda \simeq 1.30$, whereas the number of editors increases sublinearly with an exponent $\lambda \simeq 0.80$ for the later ($N > 10^6$). We use the following parameters with $10$ realizations: $b = 0.0001$, $k = 0.6$, $\tau \to \infty$, and $r = 0.01$.}  
\label{fig:model_npne_0_6}
\end{figure*}

\pagebreak
\null
\begin{figure*}[!ht]
\includegraphics[width=0.7\textwidth]{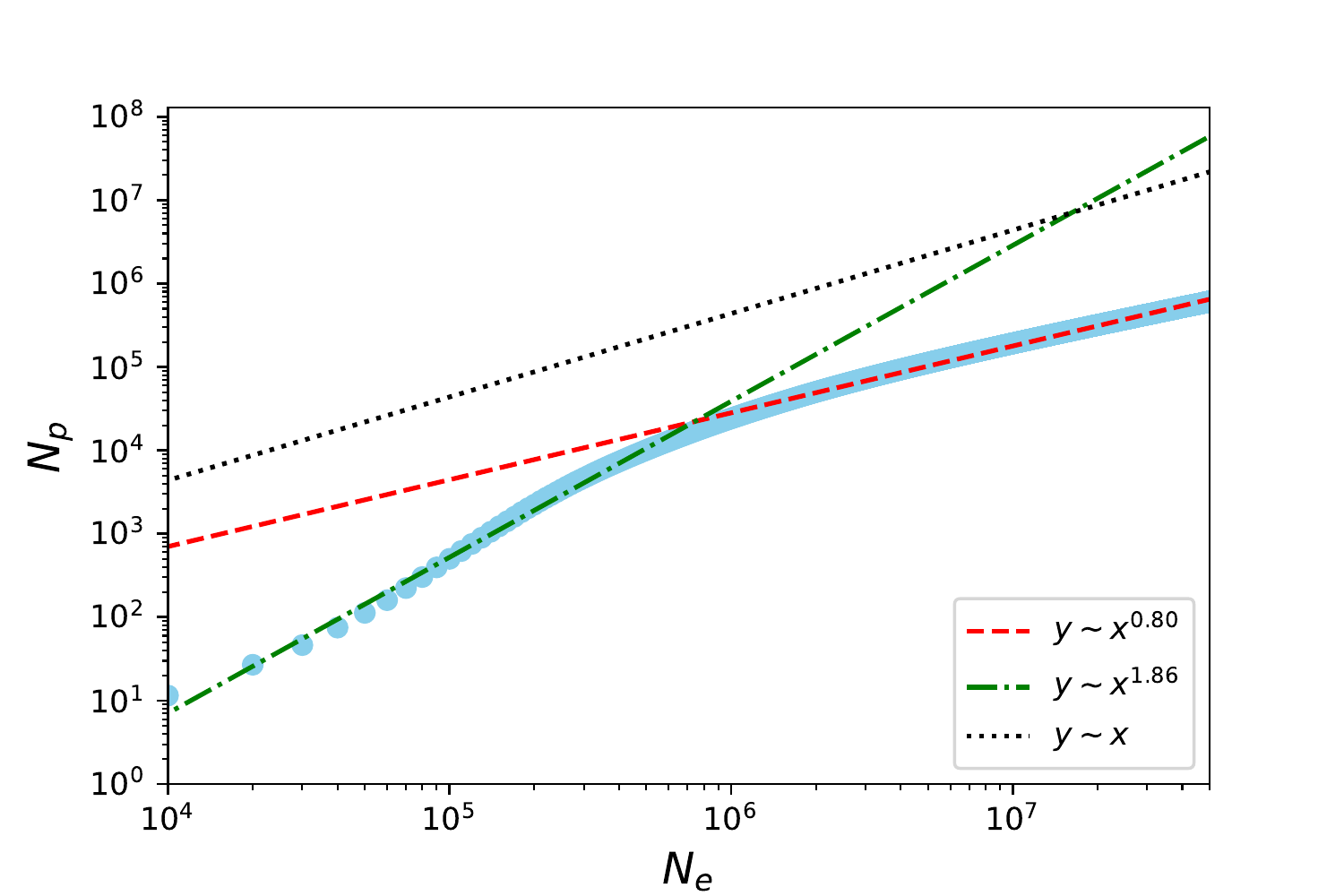}
\caption{The scaling relation between number of edits $N_e$ and number of editors $N_p$ of the model result. The relation is characterized by two simple power-law growth form of $y \sim x^\lambda$. For the initial stage $N < 5\times10^5$, the number of editors increases superlinearly with an exponent $\lambda \simeq 1.86$, whereas the number of editors increases sublinearly with an exponent $\lambda \simeq 0.80$ for the later ($N > 5\times10^5$). We use the following parameters with $10$ realizations: $b = 0.0001$, $k = 0.7$, $\tau \to \infty$, and $r = 0.01$.}  
\label{fig:model_npne_0_7}
\end{figure*}

\pagebreak
\null
\begin{figure*}[!ht]
\includegraphics[width=0.7\textwidth]{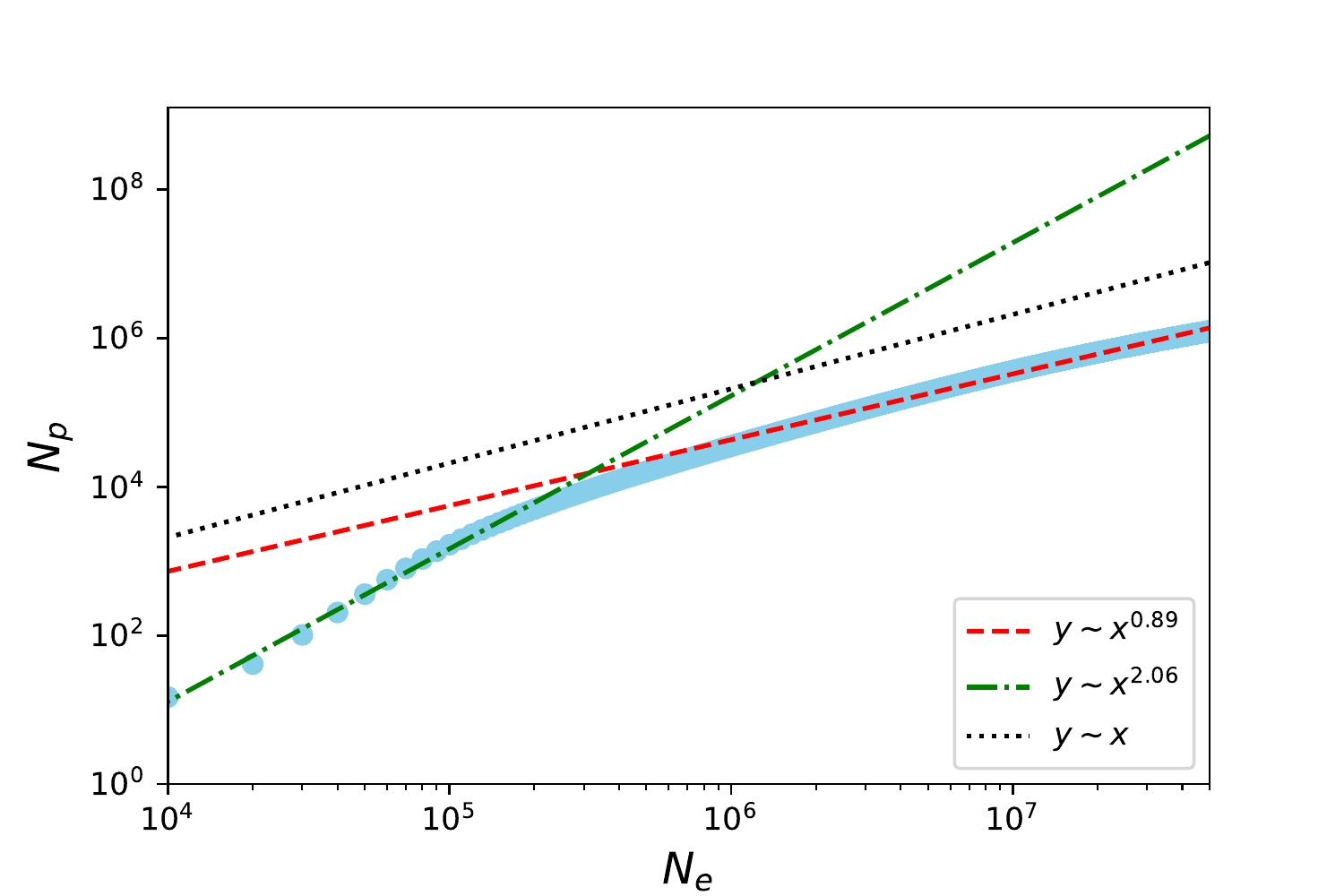}
\caption{The scaling relation between number of edits $N_e$ and number of editors $N_p$ of the model result. The relation is characterized by two simple power-law growth form of $y \sim x^\lambda$. For the initial stage $N < 2\times10^5$, the number of editors increases superlinearly with an exponent $\lambda \simeq 2.06$, whereas the number of editors increases sublinearly with an exponent $\lambda \simeq 0.89$ for the later ($N > 2\times10^5$). We use the following parameters with $10$ realizations: $b = 0.0001$, $k = 0.8$, $\tau \to \infty$, and $r = 0.01$.}  
\label{fig:model_npne_0_8}
\end{figure*}

\pagebreak
\pagebreak
\section{supplementary Tables}\label{sec:sup_tables}

\begin{table*}[!ht]
\centering
\caption{The list of Wikimedia projects. In general, there are different language editions for each project.}
\label{table01}
\begin{tabular}{lrll}
Project  &  Editions  &  Description/Notes  \\
\hline
Wikipedia  &  273  &  Encyclopedia articles   \\
Wiktionary  &  172  &  Dictionary   \\
Wikibooks  &  121  &  Educational textbooks and learning materials   \\
Wikiquote  &  89  &  Collection of quotations   \\
Wikisource  &  65  &  Library of source documents and translations  \\
Wikinews  &  33  &  News source   \\
Wikiversity  &  16  &  Educational and research materials and activities \\
Wikivoyage  &  17  &  Travel guide   \\
etc.  &  77  &    Deactivated (not editable) ones are included  \\
\hline
Total  &  863  &     \\
\end{tabular}
\end{table*}

\begin{table}[!ht]
\centering
\caption{Statistical details about power-law fitting for the $N_e$--$N_p$ relation. When we write that a $P$-value is 0, it means that this value is smaller than the minimum (approximately $2.23 \times 10^{-308}$) of the floating-point variables in Python.}
\label{table-nenp}
\begin{tabular}{lrrrrrr}
Project & \begin{tabular}[c]{@{}r@{}}Power\\     exponent\end{tabular} & \begin{tabular}[c]{@{}r@{}}Proportionality\\     coefficiant\end{tabular} & $P$-value & Standard Error & Pearson $\rho$ & \begin{tabular}[c]{@{}r@{}}Pearson $\rho$\\     (log-log)\end{tabular} \\
\hline
All & 0.704249 & 0.738648 & $0$ & 0.0110146 & 0.895056 & 0.908861 \\
Wikipedia & 0.842443 & 0.155088 & $8.82\times10^{-138}$ & 0.0192937 & 0.895667 & 0.924524 \\
Wiktionary & 0.539973 & 1.830600 & $4.72\times10^{-63}$ & 0.0201075 & 0.871989 & 0.899576 \\
Wikibooks & 0.800070 & 0.593431 & $2.52\times10^{-85}$ & 0.0148827 & 0.993453 & 0.980026 \\
Wikiquote & 0.800656 & 0.552054 & $9.37\times10^{-49}$ & 0.0258568 & 0.976789 & 0.957503 \\
Wikisource & 0.629638 & 1.278470 & $1.53\times10^{-23}$ & 0.0399435 & 0.926182 & 0.893163 \\
Wikinews & 0.622501 & 1.116610 & $2.68\times10^{-8}$ & 0.0844640 & 0.773823 & 0.797903 \\
Wikivoyage & 1.078590 & 0.017792 & $9.86\times10^{-8}$ & 0.1136140 & 0.979092 & 0.925912 \\
Wikiversity & 0.971437 & 0.068704 & $3.74\times10^{-8}$ & 0.0902719 & 0.989328 & 0.944534
\end{tabular}
\end{table}

\pagebreak

\begin{table}[!ht]
\centering
\caption{Statistical details about power-law fitting for the $N_e$--$N_a$ relation. When we write that a $P$-value is 0, it means that this value is smaller than the minimum (approximately $2.23 \times 10^{-308}$) of the floating-point variables in Python.}
\label{table-nena}
\begin{tabular}{lrrrrrr}
Project & \begin{tabular}[c]{@{}r@{}}Power\\     exponent\end{tabular} & \begin{tabular}[c]{@{}r@{}}Proportionality\\     coefficiant\end{tabular} & $P$-value & Standard Error & Pearson $\rho$ & \begin{tabular}[c]{@{}r@{}}Pearson $\rho$\\     (log-log)\end{tabular} \\
\hline
All & 0.851328 & 0.953075 & 0 & 0.0063968 & 0.851105 & 0.976546 \\
Wikipedia & 0.937205 & 0.256999 & $4.78\times10^{-217}$ & 0.0115751 & 0.849992 & 0.976168 \\
Wiktionary & 0.886418 & 0.876804 & $4.55\times10^{-133}$ & 0.0116775 & 0.971678 & 0.985567 \\
Wikibooks & 0.793356 & 1.773870 & $5.16\times10^{-104}$ & 0.0101733 & 0.979341 & 0.990358 \\
Wikiquote & 0.769828 & 1.939820 & $1.33\times10^{-75}$ & 0.0118208 & 0.949461 & 0.989899 \\
Wikisource & 0.971340 & 0.472484 & $4.70\times10^{-51}$ & 0.0204457 & 0.980636 & 0.986329 \\
Wikinews & 0.780295 & 1.999770 & $1.35\times10^{-13}$ & 0.0626789 & 0.789431 & 0.912861 \\
Wikivoyage & 0.809312 & 0.853837 & $1.04\times10^{-10}$ & 0.0516414 & 0.982997 & 0.970794 \\
Wikiversity & 0.928145 & 0.316529 & $3.56\times10^{-9}$ & 0.0717772 & 0.984704 & 0.960594
\end{tabular}
\end{table}

\begin{table}[!ht]
\centering
\caption{Statistical details about power-law fitting for the $N_e$--$S$ relation. When we write that a $P$-value is 0, it means that this value is smaller than the minimum (approximately $2.23 \times 10^{-308}$) of the floating-point variables in Python.}
\label{table-nes}
\begin{tabular}{lrrrrrr}
Project & \begin{tabular}[c]{@{}r@{}}Power\\     exponent\end{tabular} & \begin{tabular}[c]{@{}r@{}}Proportionality\\     coefficiant\end{tabular} & $P$-value & Standard Error & Pearson $\rho$ & \begin{tabular}[c]{@{}r@{}}Pearson $\rho$\\     (log-log)\end{tabular} \\
\hline
All & 0.874691 & 1060.2 & 0 & 0.0102172 & 0.945518 & 0.945977 \\
Wikipedia & 0.987816 & 194.4 & $1.68\times10^{-186}$ & 0.0153652 & 0.946319 & 0.962968 \\
Wiktionary & 0.802352 & 1652.4 & $1.59\times10^{-103}$ & 0.0160625 & 0.892158 & 0.967582 \\
Wikibooks & 0.976815 & 728.3 & $6.50\times10^{-61}$ & 0.0301235 & 0.966832 & 0.947805 \\
Wikiquote & 0.891003 & 830.3 & $1.22\times10^{-57}$ & 0.0223718 & 0.977249 & 0.973655 \\
Wikisource & 0.989188 & 1610.7 & $7.95\times10^{-34}$ & 0.0405094 & 0.951270 & 0.951021 \\
Wikinews & 0.814366 & 1697.1 & $1.79\times10^{-13}$ & 0.0661310 & 0.955948 & 0.911193 \\
Wikivoyage & 0.846150 & 1612.5 & $3.33\times10^{-7}$ & 0.0980059 & 0.956239 & 0.912402 \\
Wikiversity & 0.894525 & 1310.5 & $1.21\times10^{-8}$ & 0.0760369 & 0.981032 & 0.952962
\end{tabular}
\end{table}

\pagebreak

\begin{table}[!ht]
\centering
\caption{Statistical details about power-law fitting for the $N_a$--$N_p$ relation. When we write that a $P$-value is 0, it means that this value is smaller than the minimum (approximately $2.23 \times 10^{-308}$) of the floating-point variables in Python.}
\label{table-nanp}
\begin{tabular}{lrrrrrr}
Project & \begin{tabular}[c]{@{}r@{}}Power\\     exponent\end{tabular} & \begin{tabular}[c]{@{}r@{}}Proportionality\\     coefficiant\end{tabular} & $P$-value & Standard Error & Pearson $\rho$ & \begin{tabular}[c]{@{}r@{}}Pearson $\rho$\\     (log-log)\end{tabular} \\
\hline
All & 0.778094 & 1.199350 & $4.75\times10^{-274}$ & 0.0146429 & 0.649975 & 0.875402 \\
Wikipedia & 0.855545 & 0.817538 & $7.79\times10^{-120}$ & 0.0228277 & 0.651000 & 0.901426 \\
Wiktionary & 0.585252 & 2.473690 & $5.63\times10^{-56}$ & 0.0246021 & 0.773866 & 0.876922 \\
Wikibooks & 0.980838 & 0.403144 & $3.04\times10^{-69}$ & 0.0253558 & 0.956404 & 0.962462 \\
Wikiquote & 1.012750 & 0.338911 & $5.48\times10^{-43}$ & 0.0387238 & 0.878766 & 0.941891 \\
Wikisource & 0.616042 & 2.848200 & $4.00\times10^{-20}$ & 0.0459310 & 0.884531 & 0.860595 \\
Wikinews & 0.797727 & 0.642675 & $3.09\times10^{-11}$ & 0.0796528 & 0.955328 & 0.874015 \\
Wikivoyage & 1.188170 & 0.082902 & $1.53\times10^{-5}$ & 0.1898720 & 0.929965 & 0.850316 \\
Wikiversity & 0.940047 & 0.609241 & $5.83\times10^{-6}$ & 0.1334550 & 0.966194 & 0.883137
\end{tabular}
\end{table}

\begin{table}[!ht]
\centering
\caption{Statistical details about power-law fitting for the $N_a$--$S$ relation. When we write that a $P$-value is 0, it means that this value is smaller than the minimum (approximately $2.23 \times 10^{-308}$) of the floating-point variables in Python.}
\label{table-nas}
\begin{tabular}{lrrrrrr}
Project & \begin{tabular}[c]{@{}r@{}}Power\\     exponent\end{tabular} & \begin{tabular}[c]{@{}r@{}}Proportionality\\     coefficiant\end{tabular} & $P$-value & Standard Error & Pearson $\rho$ & \begin{tabular}[c]{@{}r@{}}Pearson $\rho$\\     (log-log)\end{tabular} \\
\hline
All & 1.021710 & 1173.3 & 0 & 0.0097043 & 0.837651 & 0.963288 \\
Wikipedia & 1.050380 & 844.92 & $6.06\times10^{-241}$ & 0.0108699 & 0.838299 & 0.983090 \\
Wiktionary & 0.900004 & 1952.1 & $1.04\times10^{-114}$ & 0.0153493 & 0.916696 & 0.976157 \\
Wikibooks & 1.231870 & 358.1 & $4.13\times10^{-66}$ & 0.0340094 & 0.958276 & 0.957518 \\
Wikiquote & 1.154880 & 392.9 & $3.45\times10^{-64}$ & 0.0241913 & 0.938157 & 0.981443 \\
Wikisource & 0.999274 & 4167.2 & $1.49\times10^{-32}$ & 0.0430899 & 0.944978 & 0.946118 \\
Wikinews & 0.981187 & 1486.8 & $7.42\times10^{-16}$ & 0.0648819 & 0.910617 & 0.938418 \\
Wikivoyage & 1.022660 & 2346.8 & $1.83\times10^{-7}$ & 0.1130380 & 0.970173 & 0.919303 \\
Wikiversity & 0.944827 & 4725.6 & $2.93\times10^{-10}$ & 0.0604193 & 0.980323 & 0.972548
\end{tabular}
\end{table}

\pagebreak

\begin{table}[!ht]
\centering
\caption{Statistical details about power-law fitting for the $N_p$--$S$ relation. When we write that a $P$-value is 0, it means that this value is smaller than the minimum (approximately $2.23 \times 10^{-308}$) of the floating-point variables in Python.}
\label{table-nps}
\begin{tabular}{lrrrrrr}
Project & \begin{tabular}[c]{@{}r@{}}Power\\     exponent\end{tabular} & \begin{tabular}[c]{@{}r@{}}Proportionality\\     coefficiant\end{tabular} & $P$-value & Standard Error & Pearson $\rho$ & \begin{tabular}[c]{@{}r@{}}Pearson $\rho$\\     (log-log)\end{tabular} \\
\hline
All & 1.059150 & 5791.2 & $4.34\times10^{-292}$ & 0.0187328 & 0.726656 & 0.887589 \\
Wikipedia & 1.022530 & 6194.5 & $1.11\times10^{-124}$ & 0.0261606 & 0.725107 & 0.908313 \\
Wiktionary & 1.223030 & 3538.5 & $2.02\times10^{-58}$ & 0.0492685 & 0.747880 & 0.885312 \\
Wikibooks & 1.160450 & 1966.0 & $5.32\times10^{-50}$ & 0.0455632 & 0.965096 & 0.919231 \\
Wikiquote & 1.029180 & 2743.8 & $1.57\times10^{-42}$ & 0.039893 & 0.967549 & 0.940423 \\
Wikisource & 1.304610 & 7218.7 & $1.68\times10^{-22}$ & 0.0868298 & 0.927855 & 0.884204 \\
Wikinews & 0.968762 & 16377.6 & $5.85\times10^{-10}$ & 0.1098120 & 0.896203 & 0.845663 \\
Wikivoyage & 0.593184 & 190739 & $5.99\times10^{-4}$ & 0.1370960 & 0.906938 & 0.745099 \\
Wikiversity & 0.809554 & 38139.5 & $4.66\times10^{-6}$ & 0.1126350 & 0.978002 & 0.887005
\end{tabular}
\end{table}

\pagebreak
\pagebreak

\section{Supplementary Methods}\label{sec:sup_methods}
\subsection*{Introduction to the Patent Data Set}\label{sec:sup_patent}
For our analysis, we use the spring 2017 edition of European Patent Office (EPO) Worldwide Patent Statistical Database (PATSTAT), which contains the metadata of patents from 91 national and international patent offices. Inventors and applicants are identified by the EPO using their person IDs and we neglected the persons who were not identified by the EPO. It contains various types of intellectual properties, but we only consider patent applications. The timestamps of patent applications are preferentially extracted from the application year field. Before 2000, the United States Patent and Trademark Office (USPTO) only published granted patents. We thus used patent applications after 2000, which was the year USPTO began to publish all patent applications. Furthermore, a certain publication is dated after 2017, which we truncated for data consistency. One should note that the person identification provided by the EPO is automatic; therefore, it is not perfect and may have unexplored errors.

\subsection*{Introduction to the Research Paper Data Set}\label{sec:sup_paper}
For our analysis on paper metadata, we use the dump of the entire SCOPUS CUSTOM XML DATA for 22 August 2017. This custom data contains the complete copy of data from the SCOPUS website from the very beginning, i.e., January 1996 to August 2017, and includes the title, journal, abstract, author information, and citation records in the XML format. Each type of document plays different roles for knowledge formation. For example, conference proceedings are a conventional method for presenting new research in the fields of computer science, whereas journal articles are the main method for many other disciplines. Some disciplines in social science also consider books and reports as important archives of knowledge. Therefore, to prevent a possible bias towards specific disciplines, we use the entire metadata regardless of the citation type in SCOPUS.

The timestamps of the publications are preferentially extracted from the \texttt{publicationdate} element. It is occasionally replaced by the \texttt{xocs:sort-year} element only if the \texttt{publicationdate} element is missing or broken. If the timestamp of a certain publication is not between 1996 and 2017, we consider the data as erroneous and remove it. In addition, users were identified as \texttt{auid} elements, along with its affiliation element of \texttt{afid}. We only count the authors with a clear identification, and the nationalities of the authors is assumed as the country to which the author's affiliation belonged at the time of publication.

\subsection*{Matching Languages of Wikimedia Projects and Their Dominant Countries}\label{sec:sup_matching}
One may ask how the current statuses of the Wikimedia projects are related to the empirical socio-economic status of each society. In Supplementary Figs.~\ref{fig:languser}--\ref{fig:paper}, we present several examples in response to the above question. An essential prerequisite is the designation of a specific language in the Wikimedia project to its dominant country. We assume that a certain language is mainly spoken in countries wherein people use the language as their official or dominant language. Therefore, we obtained language usage statistics of the countries from CIA World Factbook~\cite{CIA16} and remove the subsidiary languages to retain the official or dominant languages alone. If a country has more than three official languages, we neglect the third and lower-ranked languages based on the order of the populations that speak the languages. Subsequently, we collect the share of pageviews per Wikipedia language from a certain country based on the IP address~\cite{Wikitraffic}. Using these two data sets, we match each language used in each Wikimedia project with a specific country as follows: first, if a language is used by only one country in the CIA database, we consider the country to be substantially in possession of the Wikimedia projects written in that language. Second, if there is more than one country that uses the language as the primary language, we consider the country with the largest share of Wikipedia page view traffic as substantially occupying the Wikimedia projects written in the language. Note that the \textit{de facto} owner country of a Wikimedia project does not always match the country of origin of the language. For example, the Spanish Wikipedia is mainly used by Mexicans, and English Wikipedia is mainly used by people in the United States. We only use countries with ISO2 and ISO3 as the main country code. In total, 383 Wikimedia projects correspond to a language in the 80 distinct languages in the data set.

\subsection*{Calculating Representative Values of Socio-economic Indicators for Countries}\label{sec:sup_repval}
For capturing the socio-economic statuses of countries, we use various socio-economic indicators offered by the Central Intelligence Agency (CIA)~\cite{CIA16} and UNESCO Institute for Statistics (UIS)~\cite{UIS}. The CIA provides basic statistics such as population and the number of Internet users in different countries. UIS provides various statistics collected worldwide for four categories: 1) education and literacy, 2) science, technology, and innovation, 3) culture, and 4) communication and information. Some survey data are collected irregularly by different organisations; it is thus futile to choose a specific reference year to use. To circumvent this issue, we extract all the existing data from 2000 to 2016, and averaged them over the entire duration for a certain country and its index value. Note that the economic indicators such as Gross Domestic Product (GDP) and Gross Domestic Expenditure on R\&D (GERD) are usually collected yearly, whereas survey data such as attainment rate of school is collected irregularly. 

\subsection*{Feature Vector Construction for Clustering of Languages in Wikimedia projects}\label{sec:sup_vector}
To explore the relations between different languages in Wikimedia projects, we construct an $n$-dimensional vector for each language. Each vector element represents a characteristic measure for a specific type of Wikimedia project based on the number of its language editions (see Table \ref{table01}). Specifically, it follows the order 1) Wikipedia, 2) Wiktionary, 3) Wikibooks, 4) Wikiquote, and 5) Wikisource. As an illustrative example, the 3-vector representation of $N_e$ for English Wikimedia Projects is $[N_e(\textrm{English Wikipedia}), N_e(\textrm{English Wiktionary}), N_e(\textrm{English Wikibooks})] 
= (654\,163\,757, \\36\,453\,984, 2\,572\,276)$.

\pagebreak


\begin{thebibliography}{00}
\newcounter{firstbib}

\bibitem{Bruns2008} Bruns, A. \emph{Blogs, Wikipedia, Second Life, and beyond: From production to produsage} (Peter Lang, Bern, 2008).

\bibitem{Lemke2009} Lemke, C. \& Coughlin, E. The Change Agents. \emph{Teaching for the 21st Century} \textbf{67}, 54--59 (2009).

\bibitem{Wikipedia} \emph{Wikipedia}, Available at: \url{https://www.wikipedia.org/}.

\bibitem{Chesney2006} Chesney, T. An Empirical Examination of Wikipedia's Credibility, \emph{First Monday} \textbf{11} (2006).

\bibitem{Giles2005} Giles, J. Internet Encyclopedias Go Head to Head, \emph{Nature} \textbf{438}, 900--901 (2005).

\bibitem{Gandica2015} Gandica, Y., Carvalho, J., Sampaio dos Aidos, F. Wikipedia Editing Dynamics, \emph{Phys. Rev. E} \textbf{91} 012824 (2015).

\bibitem{Yun2016} Yun, J., Lee, S. H. \& Jeong, H. Intellectual Interchanges in the History of the Massive Online Open-editing Encyclopedia, Wikipedia, \emph{Phys. Rev. E.} \textbf{93} 012307 (2016).

\bibitem{Heaberlin2016} Heaberlin, B. \& DeDeo, S. The Evolution of Wikipedia's Norm Network, \emph{Future Internet} \textbf{8}, 14 (2016).

\bibitem{Zha2016} Zha, Y., Zhoua, T. \& Zhou, C. Unfolding large-scale online collaborative human dynamics, \emph{P. Natl. Acad. Sci. USA} \textbf{113}, 14627--14632 (2016).

\bibitem{Ortega2008} Ortega, F., Gonzalez-Barahona, J. M., \& Robles, G. On the inequality of contributions to Wikipedia, \emph{Hawaii International Conference on System Sciences, Proceedings of the 41st Annual}, 304 (2008).

\bibitem{Kittur2008} Kittur, A., Suh, B. \& Chi Ed. H. Can You Ever Trust a Wiki?: Impacting Perceived Trustworthiness in Wikipedia, \emph{CSCW '08 Proceedings of the 2008 ACM Conference on Computer Supported Cooperative Work} 477--480 (2016). 

\bibitem{Adler2008} Adler, B. T. \emph{et al.} Assigning Trust to Wikipedia Content, \emph{WikiSym '08 Proceedings of the 4th International Symposium on Wikis}, Article No. 26  (2008). 

\bibitem{Yasseri2012} Yasseri, T. \emph{et al.} Dynamics of Conflicts in Wikipedia \emph{PLOS ONE} \textbf{7}, e38869 (2012).

\bibitem{Barber1998} Barber, W. \& Badre, A. Culturability: The merging of culture and usability, \emph{Proceedings of The Fourth Conference on Human Factors and the Web} (1998).

\bibitem{Marcus2000} Marcus, A. \& Gould, E. W. Crosscurrents: Cultural dimensions and global web user-interface design, \emph{Interactions} \textbf{7}, 32--46 (2000).

\bibitem{Schmid-Isler2000} Schmid-Isler, S. The language of digital genres--A semiotic investigation of style and iconology on the World Wide Web, \emph{Proceedings of the 33rd Hawaii International Conference on System Sciences} (2000).

\bibitem{Pfeil2006} Pfeil, U., Zaphiris, P. \& Ang, C. S. Cultural Differences in Collaborative Authoring of Wikipedia, \emph{J. Comput. Mediat. Commun.} \textbf{12}, 88-113 (2006).

\bibitem{SKim2016} Kim, S. \emph{et al.} Understanding Editing Behaviors in Multilingual Wikipedia \emph{PLOS ONE} \textbf{11}, e0155305 (2016).

\bibitem{Wikimedia} \emph{Wikimedia Projects}, Available at: \url{https://wikimediafoundation.org/}.

\bibitem{WikimediaDownloads} \emph{Wikimedia Downloads}, Available at: \url{https://dumps.wikimedia.org/backup-index.html}.

\bibitem{Utf8} Yergeau, F. \emph{UTF-8, a Transformation Format of ISO 10646}, STD 63, RFC 3629 (2003).

\bibitem{Hale2014} Hale, S. A. Multilinguals and Wikipedia Editing, \emph{ACM Web Science Conference 2014} (2014).

\bibitem{WSong2003} Song, W. G., Zhang, H. P., Chen, T. \& Fan, W. C. Power-law Distribution of City Fires, \emph{Fire Safety J.} \textbf{38} 453--465 (2003).

\bibitem{Blie2006} Blei, D. M. and Jordan, M. I. Variational inference for Dirichlet process mixtures. \emph{Bayesian analysis}, \textbf{1(1)}, 121--143 (2006).

\bibitem{Maaten2008} van der Maaten, L. J. P. and Hinton, G. E. Visualizing High-Dimensional Data Using t-SNE. \emph{J. Mach. Learn. Res.} \textbf{9}, 2579--2605 (2008).

\bibitem{Ethnologue2017} \emph{Ethnologue, Summary by language size}, Available at: \url{https://www.ethnologue.com/statistics/size}.

\bibitem{Gini1912} Gini, C. \emph{Variabilita e Mutabilita (Variability and Mutability)} (C. Cuppini, Bologna, 1912).

\bibitem{Mankiw2014} Mankiw, N. G. \emph{Principles of Economics, 7th edition} (Cengage Learning, Boston, 2014).

\bibitem{George2004} George, B. P. \& George, B. P. Past Visits and the Intention to Revisit a Destination: Place Attachment as the Mediator and Novelty Seeking as the Moderator, \emph{Journal of Tourism Studies} \textbf{15}, 37--50 (2004). 

\bibitem{Crane2008} Crane, R. \& Sornette, D. Robust Dynamic Classes Revealed by Measuring the Response Function of a Social System,  \emph{P. Natl. Acad. Sci. USA} \textbf{105}, 15649--15653 (2008).

\bibitem{FWu2007} Wu, F., Huberman, B. A. Novelty and collective attention, \emph{P. Natl. Acad. Sci. USA} \textbf{104}, 17599--17601 (2007).

\bibitem{Karsai2012} Karsai, M., Kaski, K., Barab\'{a}si, A.-L. \& Kert\'{e}sz, J. Universal Features of Correlated Bursty Behaviour, \emph{Sci. Rep.} \textbf{2}, 397 (2012).

\bibitem{Karsai2014} Karsai, M., Perra, N. \& Vespignani, A. Time Varying Networks and the Weakness of Strong Ties, \emph{Sci. Rep.} \textbf{4}, 4001 (2014).

\bibitem{HJo2015} Jo, H.-H., Perotti, J. I., Kaski, K. \& Kert\'{e}sz, J. Correlated Bursts and the Role of Memory Range, \emph{Phys. Rev. E} \textbf{92}, 022814 (2015).

\bibitem{Kittur2008a} Kittur, A., \& Kraut, R. E. Harnessing the wisdom of crowds in Wikipedia: quality through coordination, \emph{Proceedings of the 2008 ACM conference on Computer supported cooperative work}, 37--46 (2008).

\bibitem{Hube2017} Hube, C. Bias in Wikipedia, \emph{Proceedings of the 26th International Conference on World Wide Web Companion}, 717--721 (2017).

\bibitem{Callahan2011} Callahan, E. S. \& Herring, S. C., Cultural bias in Wikipedia content on famous persons, \emph{J. Assoc. Inf. Sci. Technol.} \textbf{62}, 1899--1915 (2011).

\bibitem{Reagle2011} Reagle, J \& R, Lauren, Gender bias in Wikipedia and Britannica, \emph{Int. J. Commun.} \textbf{5}, 21 (2011).

\bibitem{Suh2009} Suh, B. The Singularity Is Not Near: Slowing Growth of Wikipedia, \emph{WikiSym '09 Proceedings of the 5th International Symposium on Wikis}, Article No. 8  (2009). 

\bibitem{Gherardi2013} Gherardi, M., \emph{et al.} Evidence for soft bounds in Ubuntu package sizes and mammalian body masses, \emph{P. Natl. Acad. Sci. USA} \textbf{110}, 21054--21058 (2013).

\bibitem{Tsay2014} Tsay, J., Dabbish, L., \& Herbsleb, J. Influence of social and technical factors for evaluating contribution in GitHub. \emph{Proceedings of the 36th international conference on Software engineering}, 356--366 (2014).

\bibitem{Padhye2014} Padhye, R., Mani, S., \& Sinha, V. S. A study of external community contribution to open-source projects on GitHub. \emph{In Proceedings of the 11th Working Conference on Mining Software Repositories}, 332--335(2014).

\bibitem{Benkler2006} Benkler, Y. \emph{The wealth of networks: How social production transforms markets and freedom} (Yale Univ. Press, New Haven and London, 2006).



\setcounter{firstbib}{\value{enumiv}}

\end{thebibliography}

\begin{thebibliography}{11}
\setcounter{enumiv}{\value{firstbib}}


\bibitem{Paolillo2006} Paolillo, J. C. and Anupam. D. Evaluating language statistics: The Ethnologue and beyond, \emph{Contract report for UNESCO Institute for Statistics} (2006).

\bibitem{CIA16} CIA. \emph{CIA World Factbook 2016-2017}. Available at: \url{https://www.cia.gov/library/publications/the-world-factbook/}.

\bibitem{Wikitraffic} \emph{Wikimedia Traffic Analysis Report - Page Views Per Wikipedia Language - Breakdown, Monthly requests or daily averages, for period: 1 Sep 2017 - 30 Sep 2017}, Available at: \url{https://stats.wikimedia.org/wikimedia/squids/SquidReportPageViewsPerLanguageBreakdown.htm}

\bibitem{UIS} \emph{UNESCO Institute for Statistics}, Available at: \url{http://uis.unesco.org/}.

\end{thebibliography}
\end{document}